%% file: main.tex
\documentclass[aps, prev,preprintnumbers,floatfix,nofootinbib, notitlepage, twocolumn,superscriptaddress, prr,longbibliography]{revtex4-1}

\pdfoutput=1

\usepackage{mathrsfs, amsmath, amsthm, amssymb, color, epstopdf, verbatim, enumerate, graphicx, hyperref,float}

\begin{document}

\title{Resonant enhancement of axion dark matter decay}

\begin{abstract}
The axion is particularly well motivated candidate for the dark matter comprising most of the mass of our visible Universe, leading to worldwide experimental and observational efforts towards its discovery.
As is well known, resonant cavities are a primary tool in this search, where they are used to enhance the conversion rate of axions into photons in a background electromagnetic field.
What is perhaps less well appreciated is that resonant cavities can also enhance the rate at which axions decay to two photons, a manifestation of the Purcell effect.
We examine this possibility and show that it offers a previously unexplored method to search the axion parameter space in a way that is competitive and complementary to other approaches, with the capability to probe a well-motivated region for QCD axion dark matter.
Furthermore, we establish that this method can be implemented via pre-existing axion heterodyne detection schemes, enabling such experiments to perform these searches with minimal modification.
\end{abstract}

\author{Yu-Ang Liu}
\thanks{These authors contributed equally to this work}
\author{Bilal Ahmad}
\thanks{These authors contributed equally to this work}
\author{Nick Houston}
\email{nhouston@bjut.edu.cn}
\thanks{corresponding author}
\affiliation{Institute of Theoretical Physics, School of Physics and Optoelectronic Engineering, Beijing University of Technology, Beijing 100124, China}

\maketitle{}

{\bf Introduction.}
Overwhelming evidence exists that most of the matter in our Universe is `dark', in that it interacts very feebly or not at all with the Standard Model (SM)~\cite{Rubin:1982kyu, Begeman:1991iy, Taylor:1998uk, Natarajan:2017sbo, Markevitch:2003at, Planck:2018vyg}.
In the standard $\Lambda$CDM (cold dark matter) cosmology, this dark matter (DM) is weakly interacting, non-relativistic, and cosmologically stable. 
Beyond this not much is known about the underlying nature of DM or its interactions with SM particles, which has led to a profusion of DM models and candidates.

A particularly well motivated candidate for this DM is the axion, and by extension axion-like particles \cite{Abbott:1982af, Dine:1982ah, Preskill:1982cy}.
As they arise straightforwardly from a minimal extension of the Standard Model, the Peccei-Quinn solution to the strong CP problem \cite{Peccei:1977ur, Weinberg:1977ma,Wilczek:1977pj}, axions occupy a rare focal point in theoretical physics, in that they are also a generic prediction of the exotic physics of string and M-theory compactifications \cite{Svrcek:2006yi, Arvanitaki:2009fg}.
Despite the profound differences between these two contexts, the resulting axion properties are also largely universal, presenting an easily-characterisable theoretical target.
Furthermore, as a light DM candidate coupled to the SM, axions also offer discovery potential via small scale `tabletop' experiments.

This being the case, many experiments are ongoing worldwide, seeking to discover the axion by a variety of means \cite{Irastorza:2018dyq}.
Amongst these experiments, a primary technique is the cavity haloscope, which relies upon a powerful magnetic field and a resonant cavity to enhance the rate of Primakoff conversion of axions into photons \cite{Sikivie:1983ip}.
At present, some of the most sensitive experiments searching for light DM are of this type \cite{Brubaker:2016ktl, ADMX:2018gho, Nguyen:2019xuh, HAYSTAC:2020kwv, CAPP:2020utb, Cervantes:2022epl, Cervantes:2022gtv, McAllister:2022ibe, TASEH:2022vvu, Cervantes:2022yzp, Schneemann:2023bqc, SHANHE:2023kxz, CAPP:2024dtx, He:2024fzj, APEX:2024jxw}.

In this paper, we explore an alternative channel to axion discovery via the same haloscope technology.
As we will demonstrate, in addition to enhancing the rate of Primakoff conversion of axions to single photons, resonant cavities can also be used to enhance the rate of axion decay into two photons.
This enhancement can be understood as an example of a more general phenomenon, more typically seen in the context of quantum optics, known as the Purcell effect.

We will in the following outline and explore the capability of a resonant cavity to enhance the decay rate of DM axions, before examining the resultant discovery potential of experiments making use of this principle.
Conclusions and discussion are presented in closing.

\textbf{Axion decay inside a resonant cavity.}
Following the basic principles of quantum mechanics, it is generally believed that particle decays occur spontaneously, independently of outside influence.
The Purcell effect, originally explored in Ref.~\cite{Purcell:1946}, is a rather notable counterexample to this, where the enhancement or suppression of decay processes occurs due to the environment in which these decays takes place.

We can see this effect most straightforwardly in Fermi's golden rule for transition rates, 
\begin{equation}
	\Gamma = 2\pi |\langle f| H_I |i\rangle |^2\rho\,, \quad
	\rho = dN/dE\,,
\end{equation}
where $H_I$ is an interaction Hamiltonian, and $\rho$ is the density of final states.
Altering the interaction strength in $H_I$ will affect $\Gamma$, however in principle so will modifications to $\rho$.
Whilst we expect a continuum of states in free space, inside a resonant cavity there will be a spectrum of discrete modes due to the cavity boundary conditions, which will enhance $\rho$ at certain frequencies and suppress it at others, modifying the transition rate.
This typically leads to an enhancement of the form
\begin{equation}
    \label{eq: Purcell enhancement}
    F_{\rm Purcell}\equiv\frac{\tau_{\rm free}}{\tau_{\rm cavity}}
    \simeq\frac{3\lambda^3}{4\pi^2}\left(\frac{Q}{V}\right)\,,
\end{equation}
where $\tau$ indicates the lifetime of a particular state, $\lambda$ the corresponding wavelength, $Q$ is the quality factor and $V$ the corresponding mode volume \cite{Purcell:1946}.

Turning to axion physics, from the interaction term
\begin{equation}
	\label{eq: interaction lagrangian}
	\mathcal{L} = g_{a\gamma} a \vec E\cdot \vec B\,,
\end{equation}
where $g_{a\gamma}$ is the axion-photon coupling, axions are able to convert to single photons in a background electromagnetic field, and also to decay to two photons.
In the latter case, $\tau_{\rm free}=64\pi^3/g_{a\gamma}^2m_{\rm a}^3$, where $m_{\rm a}$ is the axion mass, leading to a lifetime for typical axion DM parameters which far exceeds the age of our Universe.
To enhance this decay rate to an observable level with a cavity experiment, there are a few factors we must account for.

From Eq.~\eqref{eq: interaction lagrangian}, axion decay must produce pairs of photons with $\vec E\cdot \vec B\neq0$.
Therefore we firstly require a cavity with two resonant modes and a non-zero form factor
\begin{equation}
    \label{eq: form factor}
    \eta \equiv \frac{|\int_V\vec E\cdot\vec B|}{\sqrt{\int_V|\vec E|^2}\sqrt{\int_V |\vec B|^2}}\leq1\,.
\end{equation}
Denoting the mode frequencies by $\omega_{\rm s/p}$ (since we will subsequently identify one mode as `pump' and the other as `signal'), energy conservation dictates that $\omega_{\rm s}+ \omega_{\rm p}=\omega_{\rm a} \simeq m_{\rm a}(1+v_{\rm vir}^2/2)$, where $v_{\rm vir}$ is the virial velocity, and so in general both modes do not need to be degenerate. 
Of course, since $m_{\rm a}$ is in general unknown, we will also need to be able to scan the axion parameter space by tuning the resonant frequency of one or both modes.

Conveniently, unlike ordinary single-mode haloscope experiments this does not require a strong magnetic field, which can be a significant source of experimental complexity.
In turn, this permits the use of ultrahigh $Q$ superconducting radio frequency (SRF) cavities, which also allow for enhanced sensitivity \cite{Cervantes:2022gtv}.

Since the Purcell enhancement discussed above applies to both spontaneous and stimulated decays, we have the option to operate with an empty cavity, or with one of the modes already populated with photons.
In the latter case, photons from the so-called pump mode induce axion decays, in a manner akin to the stimulated axion decay scenario previously explored in various astrophysical settings \cite{Tkachev:1987cd, Kephart:1994uy, Rosa:2017ury, Caputo:2018vmy,Arza:2018dcy, Arza:2019nta, Arza:2021nec, Arza:2021zqc, Arza:2022dng, Arza:2023rcs}. 
One of the resulting decay photons remains in the pump mode, whilst the other appears in the signal mode, to be subsequently detected.

Pumping one cavity mode does introduce further experimental complexity, namely additional cooling requirements due to the power deposited with the cavity, and the necessity of strongly isolating pump and signal to avoid signal contamination.
However, as we will see it is nonetheless this case which offers better sensitivity.
There is also a further advantage, which we now explore.

\textbf{Axion up-conversion}
Ordinary axion haloscopes suffer from a $\sim(m_{\rm a}L)^2$ parametric suppression for small $m_{\rm a}$, where the axion Compton wavelength significantly exceeds the characteristic size of the experiment $L$.
This being the case, various authors have explored the potential of `heterodyne' or `up-conversion' detection schemes, where one cavity mode is pumped to generate an oscillating magnetic field, which acts as a background for Primakoff-type conversion but enables the aforementioned suppression to be evaded.
Low frequency axions satisfying $m_{\rm a} \simeq \omega_{\rm s} - \omega_{\rm p}$ can then be absorbed, resulting in a pump photon up-converting to a signal photon of higher energy \cite{Berlin:2019ahk, Lasenby:2019prg, Berlin:2020vrk,Li:2025pyi}, building upon the idea originally presented in Ref.~\cite{Sikivie:2010fa}.
 
Similar considerations for photon regeneration experiments have also been explored in Refs.~\cite{Janish:2019dpr, Bogorad:2019pbu, Gao:2020anb,Salnikov:2020urr}, and Refs.~\cite{Goryachev:2018vjt,Thomson:2021zvq} also employed a dual-mode strategy to search for axion-induced frequency shifts.
Cavity design requirements for such searches are analysed in Ref.~\cite{Giaccone:2022pke}.

The key point for our purposes is that axion up-conversion again requires the use of a tunable resonant cavity, with a signal and pump mode satisfying $\eta\neq 0$.
Typical implementations also rely upon ultrahigh $Q$ SRF cavities to maximise sensitivity.
Therefore, on the basis of the arguments in the previous section, this type of experiment should also be sensitive to the stimulated and spontaneous decay of axions.
Indeed, production of photon pairs via axion decay can be understood as a type of parametric down-conversion, which is a complementary process to the up-conversion outlined above.

The primary difference between these two contexts is kinematic: for axion decay (equivalently down-conversion) $m_{\rm a} \simeq \omega_{\rm s} + \omega_{\rm p}$, whilst for up-conversion $m_{\rm a} \simeq \omega_{\rm s} - \omega_{\rm p}$.
Furthermore, because the experimental conditions required for both scenarios are the same, a single experiment of this type could in principle simultaneously search for axions satisfying either condition. 

This being the case, we can leverage many results from previous research into up-conversion experiments.
However before doing so, we must calculate the signal power due to axion decay in a resonant cavity, which is unique to the current situation.

\textbf{Signal power.}
In this section we will derive the signal power due to axion decay in a resonant cavity, with full details given in the Supplemental Material~\cite{SupplementalMaterial}.
Our starting point is the axion photon interaction in Eq.~\eqref{eq: interaction lagrangian}.
Expanding in a basis of cavity modes with creation (annihilation) operators $c_{i}\,(c_{ i}^\dagger)$ for a two mode system ($i=1,2$), we find the interaction Hamiltonian
\begin{equation}
    H_{\rm int}= \frac{ig_{a\gamma}a\sqrt{\omega_{\rm 1}\omega_{\rm 2}}}{2}
    \left(\xi_{-} ({c}_1 {c}_2^{\dagger} -  {c}_1^{\dagger} {c}_2 ) + \xi_{+} ( {c}_1 {c}_2 -{c}_1^{\dagger} {c}_2^{\dagger})\right)\,,
\end{equation}
where $\omega_{i}$ is the mode energy and $\xi_\pm$ are form factors analogous to Eq.~\eqref{eq: form factor}.
We can describe our cavity system via a total Hamiltonian $H$ following the approach detailed in Ref.~\cite{PhysRevA.31.3761}, so that working in the Heisenberg picture we then have $dc_i(t)/dt = i[H,c_i]$, with a corresponding equation of motion for the down-conversion case
\begin{equation}
    \frac{dc_{\rm i}}{dt} =\left(-i \omega_i- \frac{\gamma_i}{2} \right)c_i - \sqrt{\gamma_i}b_{\rm in}- \frac{g_{a\gamma}\sqrt{\omega_{\rm s}\omega_{\rm p}}}{2} \xi_+ a c_j^\dagger \,,
\end{equation}
where $i\neq j$. 
Here $\gamma_i= \omega_i/Q_i$ and $b_{\rm in}$ capture the effects of damping and pumping respectively.

We now delineate into signal ($i=\rm s$) and pump $(j=\rm p)$ modes, Fourier transforming to solve for $c_
{\rm s}(\omega)$ we find
\begin{equation}
    \label{eq: c_s}
    c_{\rm s}(\omega) =\frac{g_{a\gamma}\sqrt{\omega_{\rm s}\omega_{\rm p}}\,\xi_+}{2}\int \frac{d\omega^{\prime}}{2\pi}\frac{a\left(\omega+\omega^{\prime}\right)c_{\,\rm p}^{\dagger}\left(\omega^{\prime}\right)}{\left(i\omega-i\omega_{\rm s}-\frac{1}{2}\gamma_{\rm s}\right)}\,,
\end{equation}
where the axion field satisfies
\begin{equation}
    \left\langle a^\dagger(\omega)a(\omega')\right\rangle = 2\pi\delta(\omega-\omega')F_{\rm a}(\omega)\rho_{\rm a}/m_{\rm a}^2\,,
\end{equation}
where $F_{\rm a}(\omega)$ is the lab-frame axion energy distribution, which follows from the underlying Maxwell-Boltzmann velocity distribution, and $\rho_{\rm a}$ is the DM density.

The expectation value of the number operator $N_s=c_s^\dagger c_s$, averaged over a sufficiently long timescale $T$, then gives the number of photons in the signal mode, 
\begin{equation}
    \left\langle N_{\rm s}\right\rangle_T 
    \equiv\frac{1}{T} \int_{-T/2}^{+T/2} dt\, \left\langle c_s(t)^\dagger c_s(t)\right\rangle\,.
\end{equation}
Inserting Eq.~\eqref{eq: c_s} we find after some manipulations
\begin{equation}
    \label{eq: signal number}
\left\langle N_{\rm s}\right\rangle_{T} \simeq\frac{g_{\,a\gamma}^{2}Q_{\rm s}\omega_{p}\left|\xi_+\right|^{2}\rho_aF_a(\omega_{\rm s}+\omega_{\rm p})}{4m_{\rm a}^2}(1 + \left\langle N_{\rm p}\right\rangle)\,.
\end{equation}
$Q_{\rm s}$ is the loaded $Q$ factor of the signal mode, where
\begin{equation}
    \label{eq: Q values}
    (Q_{\rm s})^{-1} =\left(Q_{\rm int}\right)^{-1} +\left(Q_{\rm cpl}\right)^{-1}\,,
\end{equation}
with $Q_{\rm int}$ the intrinsic $Q$ factor and $Q_{\rm cpl}$ determined by the cavity coupling to the readout.
The first term in $(1 + \left\langle N_{\rm p}\right\rangle)$ gives the spontaneous decay contribution, while $\left\langle N_{\rm p}\right\rangle$ (the average steady state number of photons in the pump mode) corresponds to stimulated decays.

Implicit in our derivation here is the assumption that $Q_{\rm s}/\omega_{\rm s}\gg Q_{\rm a}/m_{\rm a}$, where $Q_{\rm a}$ is the axion $Q$ factor, and hence that the axion power spectral density (PSD) is broader than our cavity bandwidth.
As is evident in Eq.~\eqref{eq: signal number} we therefore are only sensitive to a fraction of the total axion PSD, namely those axions which satisfy $\omega_{\rm a}\simeq \omega_{\rm s} + \omega_{\rm p}$.

As the energy in the pump mode is approximately $\omega_{\rm p}\left\langle N_{\rm p}\right\rangle$, the pump mode power escaping the cavity is $P_{\rm out}\simeq \omega_{\rm p}\left\langle N_{\rm p}\right\rangle\gamma_{\rm p}$.
Assuming a steady state the pump power into and out of the cavity is balanced, so that
\begin{equation}
    \left\langle N_{\rm p}\right\rangle
    \simeq P_{\rm in}/(\omega_{\rm p}\gamma_{\rm p})
    =P_{\rm in}Q_{\rm p}/\omega_{\rm p}^2\,.
\end{equation}
In practice $P_{\rm in}$ is fixed by the requirement that the resultant magnetic field at the cavity walls does not exceed the critical threshold to disrupt superconductivity. 
We can therefore equate ${\rm max}(P_{\rm in})$ with the dissipated power $P_{\rm diss} \simeq w_{\rm p} U_{\rm max}/Q_{\rm int}$, where the maximum stored energy $U_{\rm max}$ is determined numerically from the cavity geometry, mode profiles and material properties.

Similar logic applies to the signal power, although the $Q$-dependence is slightly different as we are interested only in the power reaching the readout rather than simply escaping the cavity.
This then yields
\begin{equation}
	\label{eq: signal power}
    P_{\rm s}\simeq 
    \omega_{\rm s}^2\left\langle N_{\rm s}\right\rangle_T/Q_{\rm cpl}\,.
\end{equation}
In practice it will be more convenient to use the corresponding PSD, which we can find by the same procedure, 
\begin{equation}
    S_{\rm s}\simeq\frac{g_{\,a\gamma}^{2}\omega_{\rm s}^3\omega_{\rm p}\,\left|\xi_+\right|^{2}\rho_{\rm a}\left(1+\left\langle N_{\rm p}\right\rangle\right)F_{\rm a}(\omega+\omega_p)}{4Q_{\rm cpl}m_{\rm a}^2\left((\omega-\omega_{\rm s})^2+\frac{1}{4}\gamma_{\rm s}^2\right)}\,.
\end{equation}

As an interesting aside, we can identify the Purcell-derived origin of $P_{\rm s}$ in the following way.
As outlined in Eq.~\eqref{eq: Purcell enhancement}, a Purcell-enhanced decay rate characteristically appears with a factor $(Q/V)$, whilst the signal power at an experiment of this type depends on this rate and the amount of DM inside the cavity, which is proportional to $V$.
Since these factors cancel the Purcell-enhanced signal power should be $V$-independent, and indeed this is precisely reflected in Eqs.~\eqref{eq: signal number} and~\eqref{eq: signal power}.
In contrast, in the ordinary haloscope case the signal power is directly proportional to $V$.

Of course, there are certain experimental considerations which curtail this `$V$-independence', most obviously that the mode frequencies are directly related to the dimensions of the cavity.
Furthermore, the necessity of cooling the cavity introduces further $V$-dependence, since cavities with a smaller surface area cannot dissipate power as effectively.

\textbf{Noise power.}
As explored in Refs.~\cite{Berlin:2019ahk, Lasenby:2019prg, Berlin:2020vrk,Li:2025pyi}, experiments of this type receive noise contributions from a variety of sources, including leakage noise, mechanical mode mixing, thermal noise, and amplifier noise.
At the higher frequencies relevant to the axion decay scenario, where $m_{\rm a} \simeq \omega_{\rm s} + \omega_{\rm p}\sim\mathcal{O}(\rm GHz)$, the primary contributions are expected to be thermal and amplifier noise.
Due to mode pumping, cooling the cavity to $\mathcal{O}({\rm K})$ temperatures is much more difficult than for ordinary haloscopes, likely requiring liquid helium cooling.
Further discussion of this point can be found in the references directly above.

Following the experimental configuration given in Ref.~\cite{Lasenby:2019prg}, we have the noise PSD
\begin{equation}
    S_{\rm n} \simeq \frac{4\beta}{(1+\beta)^2}\frac{S_{T}-S_{T_c}}{1+4Q_{\rm s}(\frac{\omega}{\omega_{\rm s}}-1)^2}+S_{T_c}+S_{\rm a}\,,
\end{equation}
where $\beta\equiv Q_{\rm int}/Q_{\rm cpl}$ is the cavity coupling parameter, $S_T=n_{T}\omega \equiv (e^{\omega/T}-1)^{-1}\,\omega$ and likewise for $S_{T_c}$, where $T_c$ is the effective temperature of back-action noise from the detector.
We assume here our system is in equilibrium at the cavity temperature, so that $T\simeq T_{\rm c}\gg\omega$, and hence $S_{T_c}\simeq S_T\simeq T$.
$S_{\rm a}$ is the amplifier noise contribution, which we assume to be subdominant to thermal noise.
Improved performance could be achieved via SQL-limited amplification, so that $S_{\rm a}=\omega$ and $T_c=0$, but we will not rely upon this here.

\textbf{Projected sensitivity.}
In this section we examine the potential sensitivity of this technique.
Since we expect the noise power in different bins to be independent, we can sum the contributions from individual bins in quadrature, giving
\begin{equation}
    \label{eq: SNR}
    {\rm SNR}^2 \simeq t_{\rm int}\int_0^\infty \frac{d\omega}{2\pi}\left(\frac{S_{\rm s}}{S_{\rm n}}\right)^2\,,
\end{equation}
with $t_{\rm int}$ is the integration time \cite{Chaudhuri:2018rqn}.
Optimal sensitivity comes from the stimulated decay case, as then $\left\langle N_{\rm p}\right\rangle\gg1$.

\begin{figure*}
  \begin{minipage}[c]{0.68\textwidth}
    \includegraphics[width=\textwidth]{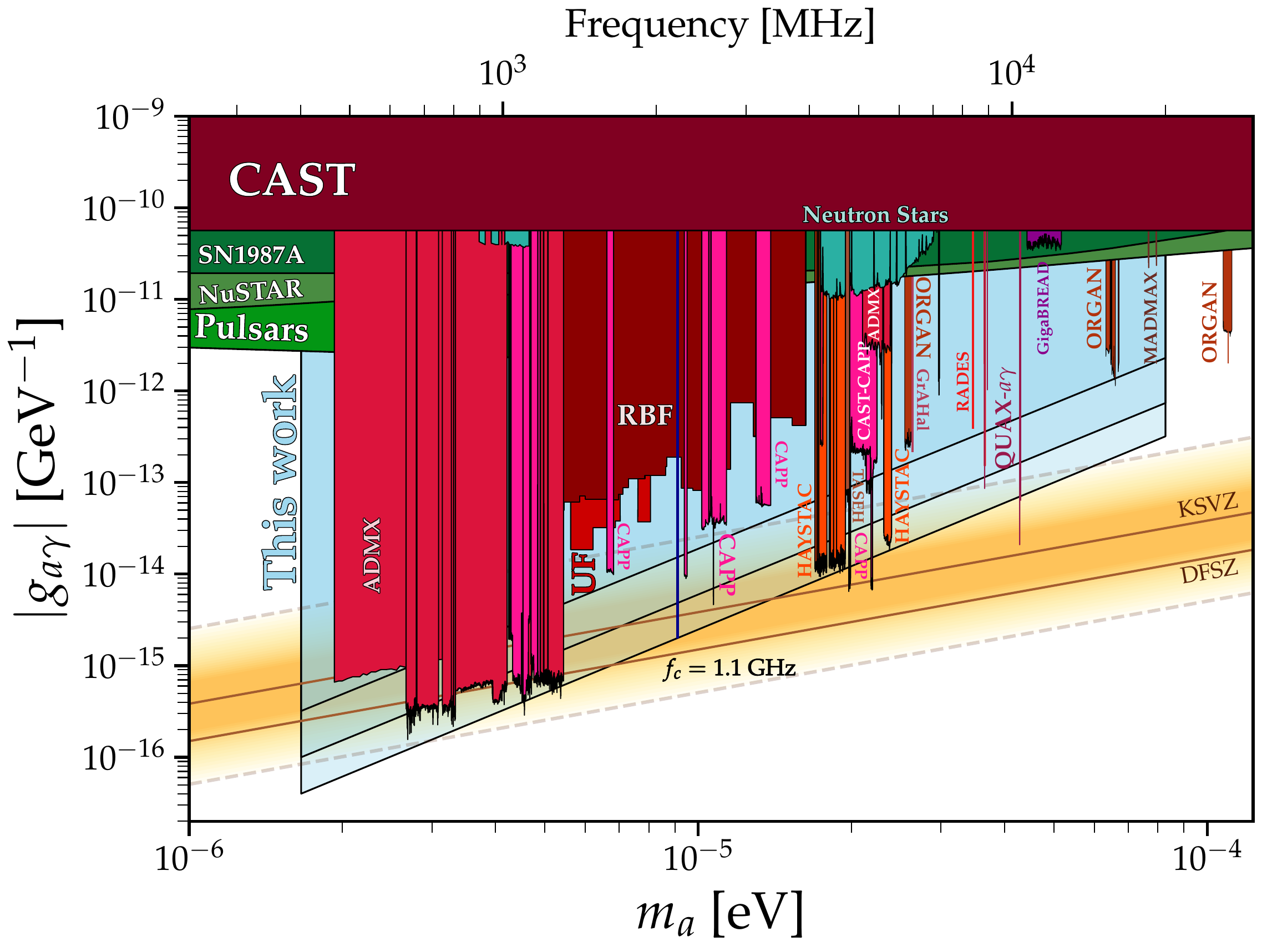}
  \end{minipage}\hfill
  \begin{minipage}[c]{0.30\textwidth}
    \label{fig: sensitivity}
    \caption{{\small
    Sensitivity for resonantly enhanced stimulated axion decay experiments based on our benchmark case, rescaling cavity dimensions to vary $f_{\rm c}$ from 0.2 to 10 GHz.
    Due to tuning range limitations multiple experiments would be required to cover this entire range, the dark blue region corresponds to a cavity exactly matching the benchmark case, as described in the text below.
    Contours shown are for ($Q_{\rm int}\simeq 2\times10^{9},\,t_{\rm int}\simeq 100\,{\rm s},\, T\simeq 2\,$K - least sensitive), ($Q_{\rm int}\simeq 2\times10^{11},\, t_{\rm int}\simeq 100\,{\rm s},\, T\simeq 2\,$K) and ($Q_{\rm int}\simeq 2\times10^{11},\,t_{\rm int}\simeq 1000\,{\rm s},\, T\simeq 1\,$K - most sensitive).
    The upper end of this range may be optimistic, since achieving such large $Q$ values at high frequency may not be feasible.
    Following Refs.~\cite{Berlin:2019ahk, Lasenby:2019prg, Berlin:2020vrk,Li:2025pyi}, we also assume sufficient pump mode rejection so as to not impact sensitivity.
    Figure production uses AxionLimits \cite{AxionLimits}, and Refs.~\cite{CAST:2017uph, Noordhuis:2022ljw, Hoof:2022xbe, Manzari:2024jns, Ruz:2024gkl, Hagmann, Hagmann:1996qd, DePanfilis, Wuensch:1989sa, ADMX:2018gho, Cervantes:2022epl, ADMX:2019uok, ADMX:2025vom, ADMX:2024pxg, Brubaker:2016ktl, HAYSTAC:2018rwy, HAYSTAC:2020kwv, HAYSTAC:2023cam, HAYSTAC:2024jch, CAPP:2020utb, CAPP:2024dtx, Kim:2022hmg, McAllister:2017lkb, Quiskamp:2022pks, Quiskamp:2023ehr, Quiskamp:2024oet, TASEH:2022vvu, Grenet:2021vbb, Adair:2022rtw, ADMX:2018ogs, Bartram:2021ysp, Hoshino:2025fiz, Alesini:2019ajt, Alesini:2020vny, Alesini:2022lnp, QUAX:2023gop, QUAX:2024fut, CAST:2020rlf, Ahyoune:2024klt, Garcia:2024xzc}.}}
  \end{minipage}
\end{figure*}
Our benchmark cavity follows Ref.~\cite{Lasenby:2019prg}, using the TE${}_{012}$ (pump) and TM${}_{013}$ (signal) modes of a cylindrical Nb cavity of length/radius $\simeq 2.35$, with the pump mode storing $U_{\rm max}\simeq690\,$J, $Q_{\rm int}\simeq2\times10^{11}$ and $V=60$ L at a frequency $f_{\rm c}=1.1$ GHz, with a form factor of 0.19.
To map to other values of $f_{\rm c}$ we simply rescale the cavity size, where $U_{\rm max}$ scales in proportion to $V$.

Our choice of this benchmark case is motivated in part by this ease of rescaling to reach other frequencies.
As previously mentioned the signal power in this case is `$V$-independent', modulo certain experimental considerations, which has the potential to evade a significant problem for cavity haloscopes, namely the diminishing sensitivity at higher frequencies due to increasingly small cavity size.
As such, it is important to assess the performance of this approach as a function of frequency.
In contrast, the cavity concepts explored in Refs.~\cite{Berlin:2019ahk, Berlin:2020vrk,Li:2025pyi} may not be as easily mapped to different frequencies.

Introducing the signal mode resonant frequency $f_{\rm s}=\omega_{\rm s}/2\pi$, Eq.~\eqref{eq: SNR} yields
\begin{align}
    \label{eq: gay limit}
    &g_{a\gamma}^{95\%}\simeq \frac{5\times10^{-15}}{{\rm GeV}}
    \left(\frac{2\times10^{11}}{Q_{\rm int}}\right)^{\frac{1}{4}} 
    \left(\frac{10^6}{Q_{\rm a}}\right)^\frac{1}{2}
    \left(\frac{f_{\rm s}}{\rm 1.3\, GHz}\right)^{\frac{9}{4}}
\nonumber\\
    &\times\left(\frac{0.45\,{\rm GeV/cm}^3}{\rho_{\rm a}}\right)^\frac{1}{2}
    \left(\frac{T}{1.8\,\rm K}\right)^{\frac{1}{2}}
    \left(\frac{100\,\rm s}{t_{\rm int}}\right)^\frac{1}{4}\,,
\end{align}
corresponding to an axion mass $m_{\rm a} \simeq 2\times \omega_{\rm s}\simeq 10.7\,\mu{\rm eV}$.
Here we assume $\omega_{\rm s}\simeq \omega_{\rm p}\simeq m_{\rm a}/2$ (bearing in mind that $\omega_{\rm s}$ and $\omega_{\rm p}$ should not be exactly equal, to enable better pump mode rejection), and that $(\omega_{\rm s}+\omega_{\rm p})$ is aligned with the peak of the axion energy distribution, so that $F_a(\omega_s+\omega_p)=60\sqrt{2\pi/e}/(17m_{\rm a}v_{\rm vir}^2)$ with $v_{\rm vir}\simeq9\times10^{-4}$ the virial velocity, where $Q_{\rm a}\simeq 1/v_{\rm vir}^2$.
$\beta$ is set to the optimal value of 2/3, and we assume all form factors are $\mathcal{O}(1)$.

Some remarks are in order regarding the tuning range of such an experiment.
In practice, tuning of SRF cavities is typically achieved via mechanical adjustment of the overall cavity length, with an $\mathcal{O}(\rm MHz)$ maximum tuning range \cite{Padamsee:1998vf}.
Alternatives to this do exist however, notably the SERAPH experiment aims to develop an SRF cavity tunable from 4 to 7 GHz for dark photon searches~\cite{Cervantes:2022gtv}.
Since we must wait for the cavity to ring up after each tuning adjustment, it preferable to use the largest possible tuning steps, namely $\mathcal{O}(m_{\rm a}/Q_{\rm a})$.
For each such step we have a sensitivity bandwidth $\Delta f=f_{\rm c}/Q_{\rm s}$.
As this is narrower than the axion PSD, we do not capture the full signal power available.
However, this can nonetheless result in better performance overall since a single tuning step can be sensitive to a range of axion mass values ~\cite{Cervantes:2022gtv}.
Taking the benchmark values above, the radiometer equation yields an instantaneous scan rate for Kim-Shifman-Vainshtein-Zakharov (KSVZ) sensitivity at $m_a\simeq 10.7\mu$eV of 1050 kHz/day.

To assess these results, we can firstly compare with the analogous expression for an ordinary single-mode Primakoff haloscope.
The world-leading constraint at $m_{\rm a}\simeq 10.7\,\mu{\rm eV}$ comes from CAPP-PACE \cite{CAPP:2020utb}, which has close to KSVZ axion sensitivity at this frequency. 
Taking their benchmark parameters we find
\begin{align}
    \label{eq: primakoff gay limit}
    g_{a\gamma}^{95\%} &\simeq \frac{10^{-14}}{\rm GeV} 
    \left(\frac{10^5}{Q}\right)^\frac{1}{2}
    \left(\frac{1.1\,\rm L}{V}\right)^\frac{1}{2} \left(\frac{7.2\,\rm T}{B_0}\right)
    \left(\frac{f}{2.6\rm\, GHz}\right)^{\frac{3}{4}}\nonumber\\
    &\times\left(\frac{0.45\,{\rm GeV/cm}^3}{\rho_{\rm a}}\right)^\frac{1}{2}
    \left(\frac{T_{\rm sys}}{1.2\,\rm K}\right)^{\frac{1}{2}}
    \left(\frac{100\,\rm s}{t_{\rm int}}\right)^\frac{1}{4}\,,
\end{align}
along with a scan rate at KSVZ sensitivity at $m_a\simeq 10.7\mu$eV of 55 kHz/day, with (optimal) $\beta=2$.

Evidently, this technique holds promise to explore a well-motivated region for QCD axion DM.
Going beyond this point of comparison, in Fig.~1 we show the sensitivity reach for various cavity sizes by rescaling our benchmark case to cover the range $f_{\rm c}=$ 0.2 to 10 GHz.
In Fig.~2 we also reproduce the sensitivity reach from Ref.~\cite{Lasenby:2019prg} to schematically show the capability of such experiments to probe axions satisfying $m_{\rm a} = \omega_{\rm s}-\omega_{\rm p}$ and $m_{\rm a} = \omega_{\rm s}+\omega_{\rm p}$.

\begin{figure}[h]
	\centering
        \includegraphics[width=1\columnwidth]{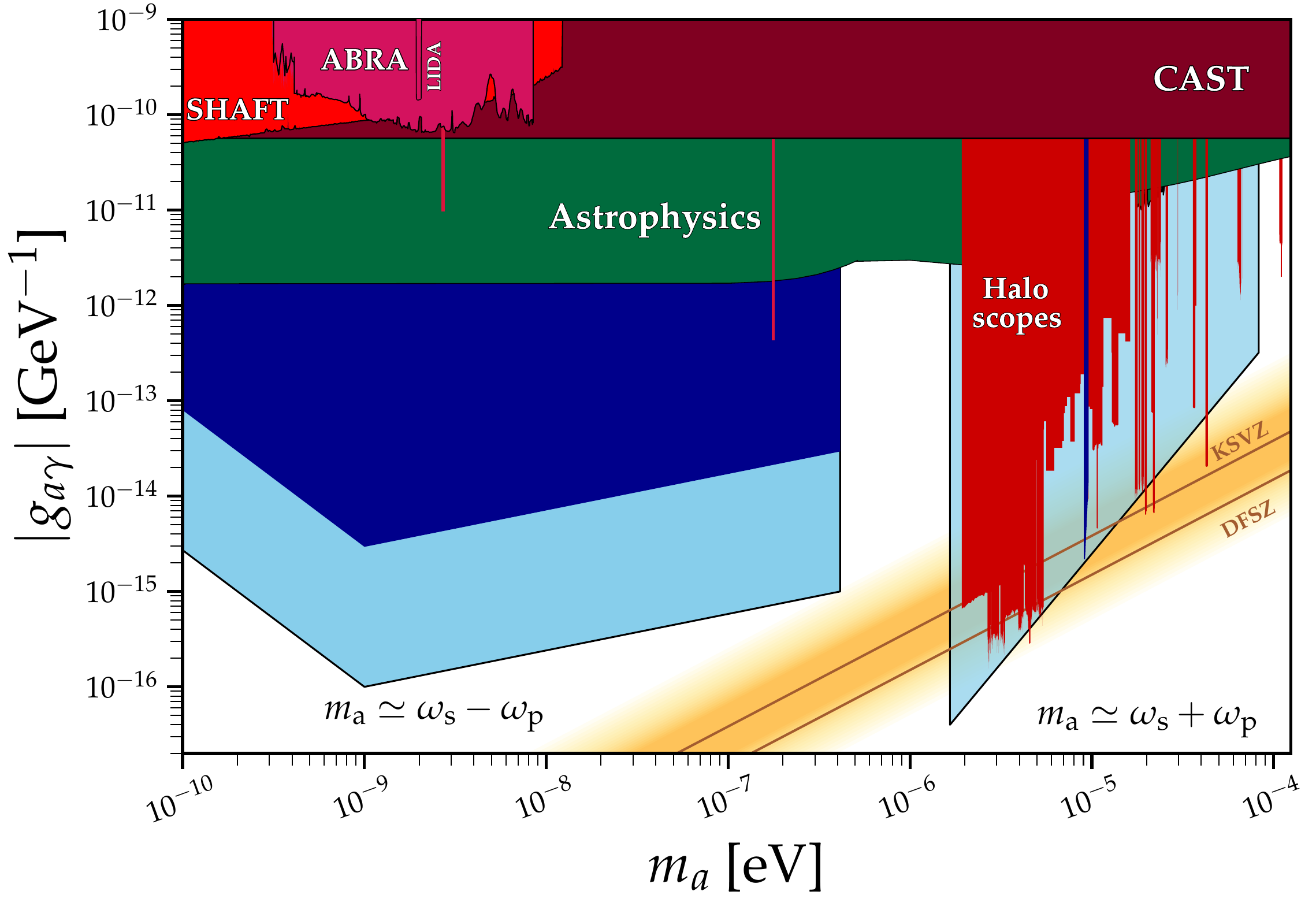}
	\caption{Sensitivity reach for experiments of this type, where the leftmost light blue region is reproduced from Ref.~\cite{Lasenby:2019prg} without modification, and the rightmost light blue region is the most sensitive contour from Fig.~1.
	Darker blue regions indicate the performance of a single cavity experiment corresponding exactly to our benchmark case at $f_{\rm c}=1.1$ GHz with the `most sensitive' parameter choices from Fig.~1, and assuming 1 hour of integration time per e-fold in axion mass values for the left hand region.
	The regions shown should be understood as somewhat heuristic, as they depend on specific details about scanning strategy and experimental configuration which are beyond the current scope of discussion.
    	As we have emphasised, a single experiment can probe parts of both regions shown, potentially simultaneously.
    	Other constraints are from Refs.~\cite{Ouellet:2018beu,Gramolin:2020ict,Salemi:2021gck,Heinze:2023nfb}.}
	\label{fig: sensitivity 2}
\end{figure}
\textbf{Discussion and conclusions.}
The axion is a particularly well-motivated candidate for the DM which makes up most of the matter in our Universe.
As a result, experimental efforts are ongoing worldwide towards its discovery.
Resonant cavities are a primary tool in this search, where they enhance the rate of axion conversion to photons in a background electromagnetic field.

In this work, we have identified and examined a previously unexplored channel for axion discovery, which also relies upon resonant cavities, by establishing that the rate of spontaneous and stimulated axion decays to two photons inside such a cavity can also be enhanced.
As we have discussed, this method allows the parameter space to be explored in a way that is competitive and complimentary to other approaches, with the capability to probe a well-motivated region for QCD axion DM.

Furthermore, we have established that SRF cavity up-conversion experiments of the type already explored in Refs.~\cite{Berlin:2019ahk, Lasenby:2019prg, Berlin:2020vrk,Li:2025pyi}, which are designed to search for low mass axions satisfying $m_{\rm a} = \omega_{\rm s}-\omega_{\rm p}$, already have the capability to search for axions satisfying $m_{\rm a} = \omega_{\rm s}+\omega_{\rm p}$ via this method.
As such, an experiment of this type could in principle search two distinct regions of the axion parameter space simultaneously.
Because these experiments are already in preparation \cite{Li:2025pyi} the scenario we explore here could also soon be directly tested, albeit potentially over a limited frequency range, owing to those experiments being optimised for up rather than down-conversion.

Going forward, these findings could be improved via further exploration and optimisation of potential experimental realisations, especially with regard to enhancing the available tuning range.
We also note that the photon pairs produced via axion decay will be polarisation entangled, potentially offering further sensitivity enhancements via quantum enhanced measurement schemes.
The applicability of this approach to `light shining through a wall' experiments is also a topic of ongoing investigation.

\textbf{Acknowledgements.} 
We thank Yifan Chen, Ariel Arza and Qiaoli Yang for useful discussions on related topics.
This work is supported by the Beijing Natural Science Foundation (Grant No. IS23025) and the National Natural Science Foundation of China (Grant No. 12150410317).

\bibliography{main}

\include{supplemental_material}

\end{document}

%% file: supplemental_material.tex
\clearpage
\newpage
\setcounter{page}{1}
\maketitle
\onecolumngrid

\begin{center}
\textbf{\large Resonant enhancement of axion dark matter decay} \\ 
\vspace{0.05in}
{ \it \large Supplemental Material}\\ 
\vspace{0.05in}
{}
{Yu-Ang Liu, Bilal Ahmad and Nick Houston}

\end{center}

\setcounter{equation}{0}
\setcounter{figure}{0}
\setcounter{table}{0}
\setcounter{section}{1}
\renewcommand{\theequation}{S\arabic{equation}}
\renewcommand{\thefigure}{S\arabic{figure}}
\renewcommand{\thetable}{S\arabic{table}}
\newcommand\ptwiddle[1]{\mathord{\mathop{#1}\limits^{\scriptscriptstyle(\sim)}}}

This supplemental material provides additional details for the calculations presented in the main body of the text.
\section{Definitions and conventions}
We work in natural units, where $\hbar=c=k_{B}=1$. 
Our Fourier convention for a function $f(t)$ is
\begin{equation}
    f\left(\omega\right)
    =\mathcal{F}\{f(t)\}
    =\int_{-\infty}^{+\infty} dt f\left(t\right)e^{i\omega t}\,,\quad
    f\left(t\right)
    =\mathcal{F}^{-1}\{f(\omega)\}
    =\frac{1}{{2\pi}}\int_{-\infty}^{+\infty} d\omega f\left(\omega\right)e^{-i\omega t}\,,
\end{equation}
where we define $f^\dagger(\omega) = (\mathcal{F}\{f(t)\})^\dagger$ rather than $\mathcal{F}\{f^\dagger(t)\}$, and similarly for $f^\dagger(t)$.
The windowed Fourier transform and power spectral density (PSD) are defined
\begin{equation}   
    f_T\left(\omega\right)
    \equiv\int_{-T/2}^{+T/2} dt f(t)e^{i\omega t}\,,\quad 
    S_{f}(\omega)\equiv\lim_{T\to\infty}\frac{1}{T}\left\langle f_T\left(\omega\right)f^\dagger_T\left(\omega\right)\right\rangle\,,
\end{equation}
where angle brackets are used to denote the ensemble average or quantum mechanical expectation value as appropriate.
Following the Wiener-Khinchin theorem, for a stationary random process we then have
\begin{equation}
    \label{eq: WK theorem}
    S_f(\omega) = \int_{-\infty}^{+\infty} dt\exp^{i\omega t}\left\langle f(t)f(0)\right\rangle\,,\quad
    \left\langle f(\omega)f^\dagger(\omega')\right\rangle = 2\pi S_f(\omega)\delta(\omega-\omega')\,.
\end{equation}

In the specific case of the axion, the energy density $\rho_{\rm a} = m_{\rm a}^2\left\langle a^2(t)\right\rangle$, while
\begin{equation}
    \label{eq: axion PSD}
    \int_{-\infty}^{+\infty} \frac{dw}{2\pi} S_{\rm a}(w) 
    =\int_{-\infty}^{+\infty} \frac{d\omega dt}{2\pi}\exp^{i\omega t}\left\langle a(t)a(0)\right\rangle
    =\left\langle a^2(t)\right\rangle\,,
\end{equation}
and so we can write $S_{\rm a}(\omega) \simeq {F}(w)\,{\rho_{\rm a}}/{m_{\rm a}^2}$ with $F$ is the corresponding axion energy distribution, which follows from the underlying Maxwell-Boltzmann velocity distribution, and satisfies $\int df\,F(f) = 1$.
In the laboratory frame
\begin{equation}
    \label{eq: axion F}
{F}(f) = 2\Theta(f-f_{\rm a}) \left( \frac{f - f_{\text{a}}}{\pi} \right)^{1/2} \left( \frac{3}{1.7 f_{\text{a}} v_{\text{vir}}^2} \right)^{3/2} \exp \left( -\frac{3(f - f_{\text{a}})}{1.7 f_{\text{a}} v_{\text{vir}}^2} \right)\,, 
\end{equation}
where $\Theta$ is the Heaviside theta function, $f_{\rm a} = m_a/2\pi$ and $v_{\rm vir} \simeq 9\times 10^{-4}$ \cite{Brubaker:2017rna}.

\section{Interaction Hamiltonian}
From $\mathcal{L} = g_{a\gamma} a\, {\bf E\cdot B}\,,$ we find the interaction Hamiltonian
\begin{equation}
    H_{int} 
    = -\int_V\mathrm{d}^3 r\, \mathcal{L}
    \simeq -g_{a \gamma } a \int_V \mathrm{d}^3 r\, \mathbf{E} \cdot \mathbf{B}\,,
\end{equation}
where spatial dependence of the axion field is neglected, and we have the following mode expansions for the electric and magnetic fields
\begin{equation}
    \mathbf{E}(t,\mathbf{r})=\sum_n\mathbf{E}_n(t,\mathbf{r}) = \sum_ni \sqrt{\frac{\omega_n}{2V_n}} (c_n - c_n^{\dagger}) \mathbf{e}_n (\mathbf{r})\,,\quad
    \mathbf{B} (t,\mathbf{r})=\sum_n\mathbf{B}_n(t,\mathbf{r}) = \sum_n \sqrt{\frac{\omega_n}{2V_n}} (c_n + c_n^\dagger) \mathbf{b}_n (\mathbf{r})\,,
\end{equation}
where $\omega_n$ is the mode energy, $V_n$ the mode volume, $c^\dagger_n$$(c_n)$ are creation (annihilation) operators, $\mathbf{e}_n/ \mathbf{b}_n$ are the spatial mode profiles, and we assume for each mode that $\mathbf{E}_i \cdot \mathbf{B}_i =0 $. 
For a two-mode system ($i=1,2$), we then find
\begin{equation}
    H_{int} = \frac{i g_{a \gamma} a}{2} \sqrt{ \frac{\omega_1 \omega_2}{V_1 V_2}}  \int_V \mathrm{d}^3 r \left((c_1 - c_1^\dagger) (c_2 + c_2^\dagger) \mathbf{e}_1 \cdot \mathbf{b}_2 + (c_2 - c_2^\dagger) (c_1 + c_1^\dagger) \mathbf{e}_2 \cdot \mathbf{b}_1\right)\,.
\end{equation}

Defining dimensionless form factors 
\begin{equation}
    \xi_1 = \frac{1}{\sqrt{V_1 V_2}} \int_V \mathrm{d}^3 r (\mathbf{e}_1 \cdot \mathbf{b}_2)\,,\quad
    \xi_2 = \frac{1}{\sqrt{V_1 V_2}} \int_V \mathrm{d}^3 r (\mathbf{e}_2 \cdot \mathbf{b}_1)\,,
\end{equation}
which encode the overlap between the two modes, we then write
\begin{equation}
    H_{int} = \frac{ig_{a \gamma} a \sqrt{\omega_1 \omega_2 }}{2}   \left(\xi_1 (c_1 - c_1^{\dagger}) (c_2 + c_2^{\dagger}) + \xi_2 (c_2 - c_2^{\dagger})(c_1 + c_1^\dagger)\right)
    = \frac{ig_{a \gamma} a \sqrt{\omega_1 \omega_2 }}{2} \left(\xi_{-} (c_1 c_2^{\dagger} - c_1^{\dagger} c_2)+\xi_{+}(c_1 c_2 - c_1^{\dagger} c_2^{\dagger})\right)\,,
\end{equation}
where $\xi_{\pm} = \xi_1 \pm \xi_2$.
The former term here ($H_u$) corresponds to up-conversion (where $\omega_{\rm a} = \omega_{i}-\omega_{j}$), while the latter ($H_d$) corresponds to down-conversion (where $\omega_{\rm a} = \omega_{i}+\omega_{j}$).

\section{Heisenberg equation of motion}
As is conventional in the context of damped quantum systems, we describe our cavity including the effects of dissipation via the input-output formalism \cite{PhysRevA.31.3761}.
Coupling the system to a thermal bath, the corresponding Hamiltonian terms are
\begin{equation}
    H_{sys} =  \sum_i\int_{-\infty}^{+\infty} \mathrm{d} \omega\, \omega\, c_i^{\dagger}(\omega) c_i(\omega)\,,\quad
    H_{bath} =  \int_{-\infty}^{+\infty} \mathrm{d} \omega\, \omega\,b^{\dagger}(\omega) b(\omega)\,,\quad
    H_{bint} = i \sum_i \int_{-\infty}^{+\infty} \mathrm{d} \omega\, \kappa_i(\omega)\left(b^{\dagger}(\omega)c_i -c_i^{\dagger} b(\omega)\right)\,,
\end{equation}
where $b^\dagger$$(b)$ are bosonic creation (annihilation) operators for the bath, which satisfy the commutation relation $[b(\omega),b^{\dagger}(\omega ')] = \delta (\omega - \omega')$.
The total Hamiltonian is then
\begin{equation}
    H = H_{sys} + H_{bath} + H_{bint} + H_{int}\,,
\end{equation}
where we can calculate for up-conversion and down-conversion separately by choosing $H_{\rm int} = H_{u}$ or $H_{d}$.

From Ref.~\cite{PhysRevA.31.3761}, in the case where $g_{a\gamma}=0$ we already have
\begin{equation}
    \frac{\mathrm{d}c_i}{\mathrm{d}t} 
    = i [H, c_i]
    = -i \omega_i c_i - \frac{\gamma_i}{2}c_i - \sqrt{\gamma_i}b_{in},
\end{equation}
where we have made the assumption that the coupling $\kappa_i(\omega) = \sqrt{{\gamma_i}/{2 \pi}}$, and introduced an `in field' to encapsulate the effect of energy transfer from the bath to the cavity.
This field satisfies 
\begin{equation}
    \label{eq: bin}
    b_{in}(t) = \frac{1}{\sqrt{2\pi}} \int \mathrm{d} \omega e^{-i \omega (t-t_0)}b_0(\omega)\,,\quad 
    \left[b_{in}(t),b^\dagger_{in}(t')\right] = \delta(t-t')\,,
\end{equation}
with $b_0(\omega)$ the value of $b(\omega)$ at $t=t_0$.

To fix the value of $\gamma_i$ we can set $b_{\rm in}=0$ and solve to find $c_i(t) = c_i(0)\exp(-i\omega_i t)\exp(-\gamma_i t/2)$.
Since intracavity power decays as the modulus squared of this amplitude, we have $P\propto\exp\left(-\gamma_i t\right)$ with decay constant 
\begin{equation}
    \gamma_i=\frac{\omega_i}{Q_i}\,.
\end{equation}

To incorporate the axion we calculate
\begin{align}
    [c_i, H_{\rm u}] 
    & = \frac{ig_{a \gamma} a \sqrt{\omega_1 \omega_2 }}{2}  \xi_{-}({c}_1[c_i,{c}_2^{\dagger}]-[c_i,{c}_1^{\dagger}]{c}_2)
    =\frac{ig_{a \gamma} a \sqrt{\omega_1 \omega_2 }}{2}  \xi_{-}\left(\delta_{i2}{c}_1-\delta_{i1}{c}_2\right)\,,\\
    [c_i, H_{\rm d}] 
    &= \frac{ig_{a \gamma} a \sqrt{\omega_1 \omega_2 }}{2} \xi_{+}  (-[c_i, {c}_1^{\dagger}] {c}_2^{\dagger}-{c}_1^{\dagger}[c_i,{c}_2^{\dagger}])
    = -\frac{ig_{a \gamma} a \sqrt{\omega_1 \omega_2 }}{2}  \xi_{+} (\delta_{i2} {c}_1^{\dagger}+\delta_{i1}{c}_2^{\dagger})\,.
\end{align}

The resulting equations of motion (for $i\neq j$) are then
\begin{align}
    \frac{\mathrm{d}c_i}{\mathrm{d}t} &= -i \omega_i c_i  - \frac{\gamma_i}{2}c_i - \sqrt{\gamma_i}b_{in}+ \frac{g_{a \gamma}\sqrt{\omega_1 \omega_2 }}{2}  \xi_{-} a\epsilon_{ij}{c}_j\,\quad \text{(up-conversion)}\,,  \\
    \frac{\mathrm{d}c_i}{\mathrm{d}t} &=-i \omega_i c_i  - \frac{\gamma_i}{2}c_i - \sqrt{\gamma_i}b_{in}-\frac{g_{a \gamma}\sqrt{\omega_1 \omega_2 }}{2} \xi_{+}a c^{\dagger}_j\,\quad\text{(down-conversion)}\,,
    \label{eq: EOM}
\end{align}
where $\epsilon_{ij}$ is the $2\times 2$ Levi-Civita symbol.

\section{Signal power}
Solving the equations of motion for $c_i(t)$ in the down-conversion case allows us to find the photon number operator $N_i = c_i^\dagger c_i$, and calculate the corresponding quantum mechanical expectation value $\left\langle N_i\right\rangle$.
Since in general this is time-dependent, we focus our attention on the steady state averaged photon number in the signal mode ($i = {\rm s}$)
\begin{equation}\label{s.567}
\left\langle N_{\rm s}\right\rangle_{T} 
\equiv\frac{1}{T}\int_{-T/2}^{+T/2}dt\,\left\langle c_{\,\rm s}^{\dagger}\left(t\right)c_{\rm s}\left(t\right)\right\rangle 
=\frac{1}{T}\int_{-T/2}^{+T/2}\int_{-\infty}^{+\infty} \frac{dt d\omega d\omega'}{(2\pi)^2} \left \langle c_{\,\rm s}^{\dagger}\left(\omega\right)c_{\rm s}\left(\omega'\right)\right\rangle e^{it(\omega-\omega')} \,,   
\end{equation}
where $T$ is some suitably long timescale.

Fourier transforming Eq.~\eqref{eq: EOM}, with no $b_{\rm in}$ contribution because we do not pump the signal mode, we find
\begin{equation}
\int_{-\infty}^{+\infty}\frac{dt d\omega^{\prime\prime}}{2\pi}\left(i\omega^{\prime\prime}-i\omega_{\rm s}-\frac{1}{2}\gamma_{\rm s}\right)c_{\rm s}\left(\omega^{\prime\prime}\right)e^{-i\left(\omega^{\prime\prime}-\omega^{\prime}\right)t}
=\frac{g_{a\gamma}\xi_+\sqrt{\omega_{\rm s}\omega_{p}}}{2}\int_{-\infty}^{+\infty}\frac{dt d\omega d\omega^{\prime\prime}}{(2\pi)^2}a\left(\omega\right)c_{\,\rm p}^{\dagger}\left(\omega^{\prime\prime}\right)e^{-i\left(\omega-\omega^{\prime\prime}-\omega^{\prime}\right)t}\,,
\end{equation}
where $c^\dagger_{\rm p}\,(c_{\rm p})$ are the creation (annihilation) operators corresponding to the pump mode ($j={\rm p}$).
Integrating over $t$ and then using the resulting $\delta$-functions, we rearrange to find
\begin{equation}
c_{\rm s}\left(\omega^{\prime}\right)=\frac{g_{a\gamma}\xi_+\sqrt{\omega_{\rm s}\omega_{\rm p}}}{2}\int_{-\infty}^{+\infty}\frac{d\omega^{\prime\prime}}{2\pi}\frac{a\left(\omega^{\prime}+\omega^{\prime\prime}\right)c_{\,\rm p}^{\dagger}\left(\omega^{\prime\prime}\right)}{\left(i\omega^{\prime}-i\omega_{\rm s}-\frac{1}{2}\gamma_{\rm s}\right)}\,,
\end{equation}
which in turn provides
\begin{equation}
\left\langle N_{\rm s}\right\rangle_{T} \simeq\frac{1}{T}\frac{g_{\,a\gamma}^{2}\omega_{\rm s}\omega_{\rm p}\,\left|\xi_+\right|^{2}}{4}\int_{-T/2}^{+T/2}\int_{-\infty}^{+\infty} \frac{dtd\omega d\omega^{\prime}d\omega^{\prime\prime}d\omega^{\prime\prime\prime}}{(2\pi)^4}\,\frac{
\left\langle a^{\dagger}\left(\omega+\omega^{\prime\prime\prime}\right)a\left(\omega'+\omega^{\prime\prime}\right)\right\rangle
\left\langle c_{\rm p}\left(\omega^{\prime\prime\prime}\right)c_{\,\rm p}^{\dagger}\left(\omega^{\prime\prime}\right)\right\rangle }
{\left(-i\omega+i\omega_{\rm s}-\frac{1}{2}\gamma_{\rm s}\right)\left(i\omega'-i\omega_{\rm s}-\frac{1}{2}\gamma_{\rm s}\right)}e^{it(\omega-\omega^\prime)}\,.  
\end{equation}  

To find $c_{\rm p}\left(\omega\right)$ we again solve Eq.~\eqref{eq: EOM}, but neglect the axion term as it is subdominant to $\sqrt{\gamma_{\rm p}}b_{\rm in}$. 
This yields
\begin{equation}
    c_{\rm p}\left(\omega\right) \simeq \frac{\sqrt{\gamma_{\rm p}}}{i\omega-i\omega_{\rm p}-\frac{1}{2}\gamma_{\rm p}}b_{\rm in}(\omega)\,,
\end{equation}
so that we then have
\begin{equation}
    \left\langle c_{\rm p}\left(\omega^{\prime\prime\prime}\right)c_{\,\rm p}^{\dagger}\left(\omega^{\prime\prime}\right)\right\rangle
    =\frac{{\gamma_{\rm p}}}{(i\omega^{\prime\prime\prime}-i\omega_{\rm p}-\frac{1}{2}\gamma_{\rm p})(-i\omega^{\prime\prime}+i\omega_{\rm p}-\frac{1}{2}\gamma_{\rm p})}\left\langle b_{\rm in}(\omega^{\prime\prime\prime})b^\dagger_{\rm in}(\omega'')\right\rangle\,.
\end{equation}
To evaluate the latter expectation value, we note that $\left\langle b_{\rm in}(\omega^{\prime\prime\prime})b^\dagger_{\rm in}(\omega'')\right\rangle=2\pi\left\langle b_{\rm 0}(\omega^{\prime\prime\prime})b^\dagger_{\rm 0}(\omega'')\right\rangle$, and in the case of a monochromatic pump mode
\begin{equation}
    \left\langle b_{0}(\omega^{\prime\prime\prime})b^\dagger_{0}(\omega^{\prime\prime})\right\rangle
    =\delta(\omega^{\prime\prime\prime}-\omega^{\prime\prime}) + \delta(\omega^{\prime\prime\prime}-\omega^{\prime\prime})\left\langle N_{\rm p}\right\rangle\,,
\end{equation}
following Ref.~\cite{PhysRevA.31.3761}.
The latter term here, with $\left\langle N_{\rm p}\right\rangle$ being the steady state average number of photons in the pump mode, corresponds to stimulated axion decay.
The former term meanwhile arises specifically due the quantum commutation relation $[b(\omega),b^\dagger(\omega^\prime)]=\delta(\omega-\omega^\prime)$, and corresponds to spontaneous axion decay.

Inserting this relation and using $\delta(\omega^{\prime\prime\prime}-\omega^{\prime\prime})$ to evaluate the $d\omega^{\prime\prime\prime}$ integral we find
\begin{equation}
\left\langle N_{\rm s}\right\rangle_{T} \simeq\frac{1}{T}\frac{g_{\,a\gamma}^{2}\omega_{\rm s}\omega_{\rm p}\,\left|\xi_+\right|^{2}(1 + \left\langle N_{\rm p}\right\rangle)}{4}\int \frac{dtd\omega d\omega^{\prime}d\omega^{\prime\prime}}{(2\pi)^3}\,\frac{
\left\langle a^{\dagger}\left(\omega+\omega^{\prime\prime}\right)a\left(\omega^\prime+\omega^{\prime\prime}\right)\right\rangle e^{it(\omega-\omega^\prime)}\gamma_p}
{\left(-i\omega+i\omega_{\rm s}-\frac{1}{2}\gamma_{\rm s}\right)\left(i\omega'-i\omega_{\rm s}-\frac{1}{2}\gamma_{\rm s}\right)\left((\omega^{\prime\prime}-\omega_{\rm p})^2+\frac{1}{4}\gamma_{\rm p}^2\right)}\,.  
\end{equation}  

From the properties of the axion PSD given in Eq.~\eqref{eq: axion PSD}, we have
\begin{equation}
    \left\langle a^{\dagger}\left(\omega+\omega^{\prime\prime}\right)a\left(\omega^\prime+\omega^{\prime\prime}\right)\right\rangle
    \simeq 2\pi\delta(\omega-\omega^\prime){F_{\rm a}(\omega+\omega^{\prime\prime})\rho_{\rm a}}/{m_{\rm a}^2}\,.
\end{equation}
Inserting this and using $\delta(\omega-\omega^\prime)$ to evaluate the $d\omega^\prime$ integral we find
\begin{equation}
\left\langle N_{\rm s}\right\rangle_{T} \simeq\frac{1}{T}\frac{g_{\,a\gamma}^{2}\omega_{\rm s}\omega_{\rm p}\,\left|\xi_+\right|^{2}(1 + \left\langle N_{\rm p}\right\rangle)\rho_{\rm a}}{4m_{\rm a}^2}\int_{-T/2}^{+T/2}\int_{-\infty}^{+\infty} \frac{dtd\omega d\omega^{\prime\prime}}{(2\pi)^2}\,\frac{
 F_{\rm a}(\omega+\omega^{\prime\prime})\gamma_p}
{\left((\omega-\omega_{\rm s})^2+\frac{1}{4}\gamma_{\rm s}^2\right)\left((\omega^{\prime\prime}-\omega_{\rm p})^2+\frac{1}{4}\gamma_{\rm p}^2\right)}\,, 
\end{equation}  
and we can now perform the $dt$ integration, which gives only a factor of $T$.

For the $d\omega^{\prime\prime}$ integration we can see the denominator has poles at $\omega^{\prime\prime} = \omega_{\rm p}\pm i\gamma_{\rm p}/2$, and so performing the residue integration and expanding for large $Q_{\rm p}$ (recalling that $\gamma_i= \omega_i/Q_i$), we then find
\begin{equation}
\label{SM: N_s}
\left\langle N_{\rm s}\right\rangle_{T} \simeq\frac{g_{\,a\gamma}^{2}\omega_{\rm s}\omega_{\rm p}\,\left|\xi_+\right|^{2}(1 + \left\langle N_{\rm p}\right\rangle)\rho_{\rm a}}{4m_{\rm a}^2}\int \frac{d\omega}{2\pi}\,\frac{F_{\rm a}(\omega+\omega_p)}
{(\omega-\omega_{\rm s})^2+\frac{1}{4}\gamma_{\rm s}^2}\,.
\end{equation}
Similarly we use residue integration to perform the integral over $d\omega$ and expand for large $Q_s$, giving
\begin{equation}
\left\langle N_{\rm s}\right\rangle_{T} \simeq\frac{g_{\,a\gamma}^{2}Q_{\rm s}\omega_{\rm p}\,\left|\xi_+\right|^{2}\rho_{\rm a}F_{\rm a}(\omega_{\rm s}+\omega_{\rm p})}{4m_{\rm a}^2}(1 + \left\langle N_{\rm p}\right\rangle)\,.
\end{equation}
Since $\left\langle N_{\rm p}\right\rangle$ can be $\gg1$, it is straightforward to see that the stimulated decay case offers the best sensitivity.

As the pump mode energy is approximately $\omega_{\rm p}\left\langle N_{\rm p}\right\rangle$, the pump mode power escaping the cavity is $P_{\rm out}\simeq \omega_{\rm p}\left\langle N_{\rm p}\right\rangle\gamma_{\rm p}$, and the pump power into and out of the cavity is balanced, so we can rearrange to find
\begin{equation}
    \left\langle N_{\rm p}\right\rangle
    \simeq \frac{P_{\rm in}}{\omega_{\rm p}\gamma_{\rm p}}
    =\frac{P_{\rm in}Q_{\rm p}}{\omega_{\rm p}^2}\,.
\end{equation}
Similar logic applies in the case of the signal mode, although we need to be careful to delineate between the  power escaping the cavity and the power actually reaching the receiver.
In the latter case $Q_{\rm cpl}$ rather than $Q_{\rm s}$ is the deciding factor, and so we then have
\begin{equation}
    \label{eq: SM P_s}
    P_{\rm s}\simeq 
    \frac{\omega_{\rm s}^2\left\langle N_{\rm s}\right\rangle_T}{Q_{\rm cpl}}
\simeq\frac{Q_{\rm s}}{Q_{\rm cpl}}
\frac{g_{\,a\gamma}^{2}\omega_{\rm s}^{2}\omega_{\rm p}\left|\xi_+\right|^{2}\rho_{\rm a}}{4m_{\rm a}^{2}}\,{F}_{\rm a}\left(\omega_{\rm s}+\omega_{\rm p}\right)\left(1+\frac{P_{\rm in}Q_{\rm p}}{\omega_{\rm p}^{2}}\right)\,.
\end{equation}
Although this expression is apparently divergent as $w_p\to0$ (since this causes $\left\langle N_{\rm p}\right\rangle$ to diverge for a fixed input power), this is not problematic because for any practical realisation of this technique the form factor $|\xi_+|^{2}$ will anyway vanish in that limit.

Going back one step, we can also use this logic to find the corresponding PSD.
Inserting the appropriate factors into Eq.~\eqref{SM: N_s} and removing the integral yields
\begin{equation}
    S_{\rm s}\left(\omega\right) \simeq\frac{\omega_{\rm s}^2}{Q_{\rm cpl}}\frac{g_{\,a\gamma}^{2}\omega_{\rm s}\omega_{\rm p}\,\left|\xi_+\right|^{2}\rho_{\rm a}}{4m_{\rm a}^2}\left(1+\frac{P_{\rm in}Q_{\rm p}}{\omega_{\rm p}^{2}}\right)\frac{F_{\rm a}(\omega+\omega_p)}
{(\omega-\omega_{\rm s})^2+\frac{1}{4}\gamma_{\rm s}^2}\,,
\end{equation}
which integrates to give Eq.~\eqref{eq: SM P_s}.

To get a feel for this result we can simplify by assuming $\omega_{\rm s}\simeq \omega_{\rm p}\simeq m_{\rm a}/2$, and introduce the normal cavity coupling parameter $\beta \equiv Q_{\rm int}/Q_{\rm cpl}$.
Of course we emphasise that $\omega_{\rm s}$ and $\omega_{\rm p}$ should not be exactly equal, to better ensure we can experimentally discriminate between signal and pump photons.
We also assume the function $F_{\rm a}\left(\omega_{\rm s}+\omega_{\rm p}\right)$ is maximised, which occurs when $\omega_{\rm s}+\omega_{\rm p}$ aligns with the peak of the axion energy distribution. 
We can find this maximum from Eq.~\eqref{eq: axion F} analytically, yielding $F_{\rm a}\left(\omega_{\rm s}+\omega_{\rm p}\right)=\text{max}\left(F_{\rm a}\right)=60\sqrt{2\pi/e}/(17m_{\rm a}v_{\rm vir}^2)$, where $e\simeq2.718$.

Introducing the resonant frequency of the signal mode $f_{\rm s}=\omega_{\rm s}/2\pi$, in this case we can then further write 
\begin{equation}
    P_{\rm s}\simeq 10^{-23}\,{\rm W}\,
    \left(\frac{g_{a\gamma}}{10^{-14}\,{\rm GeV^{-1}}}\right)^{2}
    \left(\frac{Q_{\rm p}}{10^{11}}\right)
    \left(\frac{P_{{\rm in}}}{30\,{\rm W}}\right)
    \left(\frac{\rm GHz}{f_{\rm s}}\right)^2
    \left(\frac{\beta}{1+\beta}\right)
    \left(\frac{|\xi_+|^2}{1}\right)
    \left(\frac{\rho_{{\rm a}}}{0.45\,{\rm GeV/cm^{3}}}\right)
    \left(\frac{9\times10^{-4}}{v_{\rm vir}}\right)^2\,,
\end{equation}
which corresponds to an axion mass $m_{\rm a} \simeq 2\times \omega_{\rm s}\simeq 8\,\mu{\rm eV}$.

The dependence on $v_{\rm vir}$ arises here because $Q_{\rm s}\gg Q_{\rm a}\simeq1/v_{\rm vir}^2$, so that decreasing $v_{\rm vir}$ increases the amount of DM available inside the cavity bandwidth.
This is notable because low-velocity axion flows may comprise a non-negligible component of the total DM density with velocities as low as $\mathcal{O}(10\,{\rm m/s})$ \cite{ADMX:2024pxg}, potentially leading to a significant enhancement of the corresponding signal power (since $9\times10^{-4}/10\,{\rm m/s}\simeq 10^4$).

Of course we also wish to maximise $P_{\rm in}$, which is only possible up to the extent that the resultant magnetic field at the cavity walls does not exceed the critical threshold to disrupt superconductivity. 
We can therefore equate ${\rm max}(P_{\rm in})$ with the power dissipated to the cavity walls, $P_{\rm diss} \simeq w_{\rm p} U_{\rm max}/Q_{\rm int}$, where the maximum possible stored energy $U_{\rm max}$ is determined numerically from the cavity geometry, mode profiles and material properties.

Our benchmark case here follows Ref.~\cite{Lasenby:2019prg}, which uses the TE${}_{012}$ and TM${}_{013}$ modes of a cylindrical niobium cavity of length/radius $\simeq 2.35$, with $U_{\rm max}\simeq690\,$J and $Q_{\rm int}\simeq2\times10^{11}$ at a frequency of 1.1 GHz, with a volume of 60 L and a corresponding form factor of 0.19.
To map this result to other frequencies we rescale $U_{\rm max}$ by the change in cavity volume, writing
\begin{equation}
    U_{\rm max} \simeq 690\,{\rm J}\left(\frac{V}{60\,{\rm L}}\right)\,,\quad
    V\simeq\pi r^2l\,,\quad
    r \simeq 20\,{\rm cm}\left(\frac{1.1\,{\rm GHz}}{f}\right)\,,
\end{equation}
where the length $l\simeq2.35\,r$.
Of course a more thorough numerical analysis and optimisation of cavity parameters is desirable, but this approximation suffices for our current purposes.

For this benchmark case we then find
\begin{equation}
    P_{\rm s}\simeq 10^{-23}\,{\rm W}\,
    \left(\frac{g_{a\gamma}}{10^{-14}\,{\rm GeV^{-1}}}\right)^{2}
    \left(\frac{1.1\,\rm GHz}{f_{\rm s}}\right)^4
    \left(\frac{\beta}{(1+\beta)^2}\right)
    \left(\frac{|\xi_+|^2}{1}\right)
    \left(\frac{\rho_{{\rm a}}}{0.45\,{\rm GeV/cm^{3}}}\right)
    \left(\frac{Q_{\rm a}}{10^6}\right)\,,
\end{equation}
where we use $Q_{\rm a}\simeq1/v_{\rm vir}^2$ make the $Q$-dependence explicit.
The additional factor of $1/(1+\beta)$ here relative to the Primakoff case arises because $P_{\rm in}Q_{\rm p}\propto Q_{\rm p}/Q_{\rm int}=1/(1+\beta)$, resulting in a different optimal $\beta$.

The $Q_{\rm a}$-dependence should be expected, since haloscopes ordinarily provide an enhancement factor min$(Q_{\rm a}\,,Q_{\rm s})$ \cite{Cervantes:2022gtv}, and $Q_{\rm s}\gg Q_{\rm a}$ here.
The absence of $Q_{\rm s}$-dependence meanwhile may be puzzling, however we should bear in mind that for this type of experiment the noise power $P_{\rm n}\simeq T_{\rm n}\Delta f$, where $T_{n}$ is the noise temperature and $\Delta f=f_{\rm c}/Q_{\rm s}$. 
Hence the SNR ($\sim (P_{\rm s}/P_{\rm n})\sqrt{t_{\rm int}\Delta f}\propto Q^{1/2}_{\rm s}$, where $t_{\rm int}$ is the integration time) will also be enhanced by this factor.

%% file: main.bbl
\begin{thebibliography}{97}%
\makeatletter
\providecommand \@ifxundefined [1]{%
 \@ifx{#1\undefined}
}%
\providecommand \@ifnum [1]{%
 \ifnum #1\expandafter \@firstoftwo
 \else \expandafter \@secondoftwo
 \fi
}%
\providecommand \@ifx [1]{%
 \ifx #1\expandafter \@firstoftwo
 \else \expandafter \@secondoftwo
 \fi
}%
\providecommand \natexlab [1]{#1}%
\providecommand \enquote  [1]{``#1''}%
\providecommand \bibnamefont  [1]{#1}%
\providecommand \bibfnamefont [1]{#1}%
\providecommand \citenamefont [1]{#1}%
\providecommand \href@noop [0]{\@secondoftwo}%
\providecommand \href [0]{\begingroup \@sanitize@url \@href}%
\providecommand \@href[1]{\@@startlink{#1}\@@href}%
\providecommand \@@href[1]{\endgroup#1\@@endlink}%
\providecommand \@sanitize@url [0]{\catcode `\\12\catcode `\$12\catcode
  `\&12\catcode `\#12\catcode `\^12\catcode `\_12\catcode `\%12\relax}%
\providecommand \@@startlink[1]{}%
\providecommand \@@endlink[0]{}%
\providecommand \url  [0]{\begingroup\@sanitize@url \@url }%
\providecommand \@url [1]{\endgroup\@href {#1}{\urlprefix }}%
\providecommand \urlprefix  [0]{URL }%
\providecommand \Eprint [0]{\href }%
\providecommand \doibase [0]{http://dx.doi.org/}%
\providecommand \selectlanguage [0]{\@gobble}%
\providecommand \bibinfo  [0]{\@secondoftwo}%
\providecommand \bibfield  [0]{\@secondoftwo}%
\providecommand \translation [1]{[#1]}%
\providecommand \BibitemOpen [0]{}%
\providecommand \bibitemStop [0]{}%
\providecommand \bibitemNoStop [0]{.\EOS\space}%
\providecommand \EOS [0]{\spacefactor3000\relax}%
\providecommand \BibitemShut  [1]{\csname bibitem#1\endcsname}%
\let\auto@bib@innerbib\@empty
\bibitem [{\citenamefont {Rubin}\ \emph {et~al.}(1982)\citenamefont {Rubin},
  \citenamefont {Ford}, \citenamefont {Thonnard},\ and\ \citenamefont
  {Burstein}}]{Rubin:1982kyu}%
  \BibitemOpen
  \bibfield  {author} {\bibinfo {author} {\bibfnamefont {Vera~C.}\ \bibnamefont
  {Rubin}}, \bibinfo {author} {\bibfnamefont {W.~Kent}\ \bibnamefont {Ford},
  \bibfnamefont {Jr.}}, \bibinfo {author} {\bibfnamefont {Norbert}\
  \bibnamefont {Thonnard}}, \ and\ \bibinfo {author} {\bibfnamefont {David}\
  \bibnamefont {Burstein}},\ }\bibfield  {title} {\enquote {\bibinfo {title}
  {{Rotational properties of 23 SB galaxies}},}\ }\href {\doibase
  10.1086/160355} {\bibfield  {journal} {\bibinfo  {journal} {Astrophys. J.}\
  }\textbf {\bibinfo {volume} {261}},\ \bibinfo {pages} {439} (\bibinfo {year}
  {1982})}\BibitemShut {NoStop}%
\bibitem [{\citenamefont {Begeman}\ \emph {et~al.}(1991)\citenamefont
  {Begeman}, \citenamefont {Broeils},\ and\ \citenamefont
  {Sanders}}]{Begeman:1991iy}%
  \BibitemOpen
  \bibfield  {author} {\bibinfo {author} {\bibfnamefont {K.~G.}\ \bibnamefont
  {Begeman}}, \bibinfo {author} {\bibfnamefont {A.~H.}\ \bibnamefont
  {Broeils}}, \ and\ \bibinfo {author} {\bibfnamefont {R.~H.}\ \bibnamefont
  {Sanders}},\ }\bibfield  {title} {\enquote {\bibinfo {title} {{Extended
  rotation curves of spiral galaxies: Dark haloes and modified dynamics}},}\
  }\href {\doibase 10.1093/mnras/249.3.523} {\bibfield  {journal} {\bibinfo
  {journal} {Mon. Not. Roy. Astron. Soc.}\ }\textbf {\bibinfo {volume} {249}},\
  \bibinfo {pages} {523} (\bibinfo {year} {1991})}\BibitemShut {NoStop}%
\bibitem [{\citenamefont {Taylor}\ \emph {et~al.}(1998)\citenamefont {Taylor},
  \citenamefont {Dye}, \citenamefont {Broadhurst}, \citenamefont {Benitez},\
  and\ \citenamefont {van Kampen}}]{Taylor:1998uk}%
  \BibitemOpen
  \bibfield  {author} {\bibinfo {author} {\bibfnamefont {A.~N.}\ \bibnamefont
  {Taylor}}, \bibinfo {author} {\bibfnamefont {S.}~\bibnamefont {Dye}},
  \bibinfo {author} {\bibfnamefont {Thomas~J.}\ \bibnamefont {Broadhurst}},
  \bibinfo {author} {\bibfnamefont {N.}~\bibnamefont {Benitez}}, \ and\
  \bibinfo {author} {\bibfnamefont {E.}~\bibnamefont {van Kampen}},\ }\bibfield
   {title} {\enquote {\bibinfo {title} {{Gravitational lens magnification and
  the mass of Abell 1689}},}\ }\href {\doibase 10.1086/305827} {\bibfield
  {journal} {\bibinfo  {journal} {Astrophys. J.}\ }\textbf {\bibinfo {volume}
  {501}},\ \bibinfo {pages} {539} (\bibinfo {year} {1998})},\ \Eprint
  {http://arxiv.org/abs/astro-ph/9801158} {arXiv:astro-ph/9801158} \BibitemShut
  {NoStop}%
\bibitem [{\citenamefont {Natarajan}\ \emph {et~al.}(2017)\citenamefont
  {Natarajan} \emph {et~al.}}]{Natarajan:2017sbo}%
  \BibitemOpen
  \bibfield  {author} {\bibinfo {author} {\bibfnamefont {Priyamvada}\
  \bibnamefont {Natarajan}} \emph {et~al.},\ }\bibfield  {title} {\enquote
  {\bibinfo {title} {{Mapping substructure in the HST Frontier Fields cluster
  lenses and in cosmological simulations}},}\ }\href {\doibase
  10.1093/mnras/stw3385} {\bibfield  {journal} {\bibinfo  {journal} {Mon. Not.
  Roy. Astron. Soc.}\ }\textbf {\bibinfo {volume} {468}},\ \bibinfo {pages}
  {1962--1980} (\bibinfo {year} {2017})},\ \Eprint
  {http://arxiv.org/abs/1702.04348} {arXiv:1702.04348 [astro-ph.GA]}
  \BibitemShut {NoStop}%
\bibitem [{\citenamefont {Markevitch}\ \emph {et~al.}(2004)\citenamefont
  {Markevitch}, \citenamefont {Gonzalez}, \citenamefont {Clowe}, \citenamefont
  {Vikhlinin}, \citenamefont {David}, \citenamefont {Forman}, \citenamefont
  {Jones}, \citenamefont {Murray},\ and\ \citenamefont
  {Tucker}}]{Markevitch:2003at}%
  \BibitemOpen
  \bibfield  {author} {\bibinfo {author} {\bibfnamefont {Maxim}\ \bibnamefont
  {Markevitch}}, \bibinfo {author} {\bibfnamefont {A.~H.}\ \bibnamefont
  {Gonzalez}}, \bibinfo {author} {\bibfnamefont {D.}~\bibnamefont {Clowe}},
  \bibinfo {author} {\bibfnamefont {A.}~\bibnamefont {Vikhlinin}}, \bibinfo
  {author} {\bibfnamefont {L.}~\bibnamefont {David}}, \bibinfo {author}
  {\bibfnamefont {W.}~\bibnamefont {Forman}}, \bibinfo {author} {\bibfnamefont
  {C.}~\bibnamefont {Jones}}, \bibinfo {author} {\bibfnamefont
  {S.}~\bibnamefont {Murray}}, \ and\ \bibinfo {author} {\bibfnamefont
  {W.}~\bibnamefont {Tucker}},\ }\bibfield  {title} {\enquote {\bibinfo {title}
  {{Direct constraints on the dark matter self-interaction cross-section from
  the merging galaxy cluster 1E0657-56}},}\ }\href {\doibase 10.1086/383178}
  {\bibfield  {journal} {\bibinfo  {journal} {Astrophys. J.}\ }\textbf
  {\bibinfo {volume} {606}},\ \bibinfo {pages} {819--824} (\bibinfo {year}
  {2004})},\ \Eprint {http://arxiv.org/abs/astro-ph/0309303}
  {arXiv:astro-ph/0309303} \BibitemShut {NoStop}%
\bibitem [{\citenamefont {Aghanim}\ \emph {et~al.}(2020)\citenamefont {Aghanim}
  \emph {et~al.}}]{Planck:2018vyg}%
  \BibitemOpen
  \bibfield  {author} {\bibinfo {author} {\bibfnamefont {N.}~\bibnamefont
  {Aghanim}} \emph {et~al.} (\bibinfo {collaboration} {Planck}),\ }\bibfield
  {title} {\enquote {\bibinfo {title} {{Planck 2018 results. VI. Cosmological
  parameters}},}\ }\href {\doibase 10.1051/0004-6361/201833910} {\bibfield
  {journal} {\bibinfo  {journal} {Astron. Astrophys.}\ }\textbf {\bibinfo
  {volume} {641}},\ \bibinfo {pages} {A6} (\bibinfo {year} {2020})},\ \bibinfo
  {note} {[Erratum: Astron.Astrophys. 652, C4 (2021)]},\ \Eprint
  {http://arxiv.org/abs/1807.06209} {arXiv:1807.06209 [astro-ph.CO]}
  \BibitemShut {NoStop}%
\bibitem [{\citenamefont {Abbott}\ and\ \citenamefont
  {Sikivie}(1983)}]{Abbott:1982af}%
  \BibitemOpen
  \bibfield  {author} {\bibinfo {author} {\bibfnamefont {L.~F.}\ \bibnamefont
  {Abbott}}\ and\ \bibinfo {author} {\bibfnamefont {P.}~\bibnamefont
  {Sikivie}},\ }\bibfield  {title} {\enquote {\bibinfo {title} {{A Cosmological
  Bound on the Invisible Axion}},}\ }\href {\doibase
  10.1016/0370-2693(83)90638-X} {\bibfield  {journal} {\bibinfo  {journal}
  {Phys. Lett. B}\ }\textbf {\bibinfo {volume} {120}},\ \bibinfo {pages}
  {133--136} (\bibinfo {year} {1983})}\BibitemShut {NoStop}%
\bibitem [{\citenamefont {Dine}\ and\ \citenamefont
  {Fischler}(1983)}]{Dine:1982ah}%
  \BibitemOpen
  \bibfield  {author} {\bibinfo {author} {\bibfnamefont {Michael}\ \bibnamefont
  {Dine}}\ and\ \bibinfo {author} {\bibfnamefont {Willy}\ \bibnamefont
  {Fischler}},\ }\bibfield  {title} {\enquote {\bibinfo {title} {{The Not So
  Harmless Axion}},}\ }\href {\doibase 10.1016/0370-2693(83)90639-1} {\bibfield
   {journal} {\bibinfo  {journal} {Phys. Lett. B}\ }\textbf {\bibinfo {volume}
  {120}},\ \bibinfo {pages} {137--141} (\bibinfo {year} {1983})}\BibitemShut
  {NoStop}%
\bibitem [{\citenamefont {Preskill}\ \emph {et~al.}(1983)\citenamefont
  {Preskill}, \citenamefont {Wise},\ and\ \citenamefont
  {Wilczek}}]{Preskill:1982cy}%
  \BibitemOpen
  \bibfield  {author} {\bibinfo {author} {\bibfnamefont {John}\ \bibnamefont
  {Preskill}}, \bibinfo {author} {\bibfnamefont {Mark~B.}\ \bibnamefont
  {Wise}}, \ and\ \bibinfo {author} {\bibfnamefont {Frank}\ \bibnamefont
  {Wilczek}},\ }\bibfield  {title} {\enquote {\bibinfo {title} {{Cosmology of
  the Invisible Axion}},}\ }\href {\doibase 10.1016/0370-2693(83)90637-8}
  {\bibfield  {journal} {\bibinfo  {journal} {Phys. Lett. B}\ }\textbf
  {\bibinfo {volume} {120}},\ \bibinfo {pages} {127--132} (\bibinfo {year}
  {1983})}\BibitemShut {NoStop}%
\bibitem [{\citenamefont {Peccei}\ and\ \citenamefont
  {Quinn}(1977)}]{Peccei:1977ur}%
  \BibitemOpen
  \bibfield  {author} {\bibinfo {author} {\bibfnamefont {R.~D.}\ \bibnamefont
  {Peccei}}\ and\ \bibinfo {author} {\bibfnamefont {Helen~R.}\ \bibnamefont
  {Quinn}},\ }\bibfield  {title} {\enquote {\bibinfo {title} {{Constraints
  Imposed by CP Conservation in the Presence of Instantons}},}\ }\href
  {\doibase 10.1103/PhysRevD.16.1791} {\bibfield  {journal} {\bibinfo
  {journal} {Phys. Rev. D}\ }\textbf {\bibinfo {volume} {16}},\ \bibinfo
  {pages} {1791--1797} (\bibinfo {year} {1977})}\BibitemShut {NoStop}%
\bibitem [{\citenamefont {Weinberg}(1978)}]{Weinberg:1977ma}%
  \BibitemOpen
  \bibfield  {author} {\bibinfo {author} {\bibfnamefont {Steven}\ \bibnamefont
  {Weinberg}},\ }\bibfield  {title} {\enquote {\bibinfo {title} {{A New Light
  Boson?}}}\ }\href {\doibase 10.1103/PhysRevLett.40.223} {\bibfield  {journal}
  {\bibinfo  {journal} {Phys. Rev. Lett.}\ }\textbf {\bibinfo {volume} {40}},\
  \bibinfo {pages} {223--226} (\bibinfo {year} {1978})}\BibitemShut {NoStop}%
\bibitem [{\citenamefont {Wilczek}(1978)}]{Wilczek:1977pj}%
  \BibitemOpen
  \bibfield  {author} {\bibinfo {author} {\bibfnamefont {Frank}\ \bibnamefont
  {Wilczek}},\ }\bibfield  {title} {\enquote {\bibinfo {title} {{Problem of
  Strong $P$ and $T$ Invariance in the Presence of Instantons}},}\ }\href
  {\doibase 10.1103/PhysRevLett.40.279} {\bibfield  {journal} {\bibinfo
  {journal} {Phys. Rev. Lett.}\ }\textbf {\bibinfo {volume} {40}},\ \bibinfo
  {pages} {279--282} (\bibinfo {year} {1978})}\BibitemShut {NoStop}%
\bibitem [{\citenamefont {Svrcek}\ and\ \citenamefont
  {Witten}(2006)}]{Svrcek:2006yi}%
  \BibitemOpen
  \bibfield  {author} {\bibinfo {author} {\bibfnamefont {Peter}\ \bibnamefont
  {Svrcek}}\ and\ \bibinfo {author} {\bibfnamefont {Edward}\ \bibnamefont
  {Witten}},\ }\bibfield  {title} {\enquote {\bibinfo {title} {{Axions In
  String Theory}},}\ }\href {\doibase 10.1088/1126-6708/2006/06/051} {\bibfield
   {journal} {\bibinfo  {journal} {JHEP}\ }\textbf {\bibinfo {volume} {06}},\
  \bibinfo {pages} {051} (\bibinfo {year} {2006})},\ \Eprint
  {http://arxiv.org/abs/hep-th/0605206} {arXiv:hep-th/0605206} \BibitemShut
  {NoStop}%
\bibitem [{\citenamefont {Arvanitaki}\ \emph {et~al.}(2010)\citenamefont
  {Arvanitaki}, \citenamefont {Dimopoulos}, \citenamefont {Dubovsky},
  \citenamefont {Kaloper},\ and\ \citenamefont
  {March-Russell}}]{Arvanitaki:2009fg}%
  \BibitemOpen
  \bibfield  {author} {\bibinfo {author} {\bibfnamefont {Asimina}\ \bibnamefont
  {Arvanitaki}}, \bibinfo {author} {\bibfnamefont {Savas}\ \bibnamefont
  {Dimopoulos}}, \bibinfo {author} {\bibfnamefont {Sergei}\ \bibnamefont
  {Dubovsky}}, \bibinfo {author} {\bibfnamefont {Nemanja}\ \bibnamefont
  {Kaloper}}, \ and\ \bibinfo {author} {\bibfnamefont {John}\ \bibnamefont
  {March-Russell}},\ }\bibfield  {title} {\enquote {\bibinfo {title} {{String
  Axiverse}},}\ }\href {\doibase 10.1103/PhysRevD.81.123530} {\bibfield
  {journal} {\bibinfo  {journal} {Phys. Rev. D}\ }\textbf {\bibinfo {volume}
  {81}},\ \bibinfo {pages} {123530} (\bibinfo {year} {2010})},\ \Eprint
  {http://arxiv.org/abs/0905.4720} {arXiv:0905.4720 [hep-th]} \BibitemShut
  {NoStop}%
\bibitem [{\citenamefont {Irastorza}\ and\ \citenamefont
  {Redondo}(2018)}]{Irastorza:2018dyq}%
  \BibitemOpen
  \bibfield  {author} {\bibinfo {author} {\bibfnamefont {Igor~G.}\ \bibnamefont
  {Irastorza}}\ and\ \bibinfo {author} {\bibfnamefont {Javier}\ \bibnamefont
  {Redondo}},\ }\bibfield  {title} {\enquote {\bibinfo {title} {{New
  experimental approaches in the search for axion-like particles}},}\ }\href
  {\doibase 10.1016/j.ppnp.2018.05.003} {\bibfield  {journal} {\bibinfo
  {journal} {Prog. Part. Nucl. Phys.}\ }\textbf {\bibinfo {volume} {102}},\
  \bibinfo {pages} {89--159} (\bibinfo {year} {2018})},\ \Eprint
  {http://arxiv.org/abs/1801.08127} {arXiv:1801.08127 [hep-ph]} \BibitemShut
  {NoStop}%
\bibitem [{\citenamefont {Sikivie}(1983)}]{Sikivie:1983ip}%
  \BibitemOpen
  \bibfield  {author} {\bibinfo {author} {\bibfnamefont {P.}~\bibnamefont
  {Sikivie}},\ }\bibfield  {title} {\enquote {\bibinfo {title} {{Experimental
  Tests of the Invisible Axion}},}\ }\href {\doibase
  10.1103/PhysRevLett.51.1415} {\bibfield  {journal} {\bibinfo  {journal}
  {Phys. Rev. Lett.}\ }\textbf {\bibinfo {volume} {51}},\ \bibinfo {pages}
  {1415--1417} (\bibinfo {year} {1983})},\ \bibinfo {note} {[Erratum:
  Phys.Rev.Lett. 52, 695 (1984)]}\BibitemShut {NoStop}%
\bibitem [{\citenamefont {Brubaker}\ \emph
  {et~al.}(2017{\natexlab{a}})\citenamefont {Brubaker} \emph
  {et~al.}}]{Brubaker:2016ktl}%
  \BibitemOpen
  \bibfield  {author} {\bibinfo {author} {\bibfnamefont {B.~M.}\ \bibnamefont
  {Brubaker}} \emph {et~al.},\ }\bibfield  {title} {\enquote {\bibinfo {title}
  {{First results from a microwave cavity axion search at 24 $\mu$eV}},}\
  }\href {\doibase 10.1103/PhysRevLett.118.061302} {\bibfield  {journal}
  {\bibinfo  {journal} {Phys. Rev. Lett.}\ }\textbf {\bibinfo {volume} {118}},\
  \bibinfo {pages} {061302} (\bibinfo {year} {2017}{\natexlab{a}})},\ \Eprint
  {http://arxiv.org/abs/1610.02580} {arXiv:1610.02580 [astro-ph.CO]}
  \BibitemShut {NoStop}%
\bibitem [{\citenamefont {Du}\ \emph {et~al.}(2018)\citenamefont {Du} \emph
  {et~al.}}]{ADMX:2018gho}%
  \BibitemOpen
  \bibfield  {author} {\bibinfo {author} {\bibfnamefont {N.}~\bibnamefont {Du}}
  \emph {et~al.} (\bibinfo {collaboration} {ADMX}),\ }\bibfield  {title}
  {\enquote {\bibinfo {title} {{A Search for Invisible Axion Dark Matter with
  the Axion Dark Matter Experiment}},}\ }\href {\doibase
  10.1103/PhysRevLett.120.151301} {\bibfield  {journal} {\bibinfo  {journal}
  {Phys. Rev. Lett.}\ }\textbf {\bibinfo {volume} {120}},\ \bibinfo {pages}
  {151301} (\bibinfo {year} {2018})},\ \Eprint
  {http://arxiv.org/abs/1804.05750} {arXiv:1804.05750 [hep-ex]} \BibitemShut
  {NoStop}%
\bibitem [{\citenamefont {Nguyen}\ \emph {et~al.}(2019)\citenamefont {Nguyen},
  \citenamefont {Lobanov},\ and\ \citenamefont {Horns}}]{Nguyen:2019xuh}%
  \BibitemOpen
  \bibfield  {author} {\bibinfo {author} {\bibfnamefont {Le~Hoang}\
  \bibnamefont {Nguyen}}, \bibinfo {author} {\bibfnamefont {Andrei}\
  \bibnamefont {Lobanov}}, \ and\ \bibinfo {author} {\bibfnamefont {Dieter}\
  \bibnamefont {Horns}},\ }\bibfield  {title} {\enquote {\bibinfo {title}
  {{First results from the WISPDMX radio frequency cavity searches for hidden
  photon dark matter}},}\ }\href {\doibase 10.1088/1475-7516/2019/10/014}
  {\bibfield  {journal} {\bibinfo  {journal} {JCAP}\ }\textbf {\bibinfo
  {volume} {10}},\ \bibinfo {pages} {014} (\bibinfo {year} {2019})},\ \Eprint
  {http://arxiv.org/abs/1907.12449} {arXiv:1907.12449 [hep-ex]} \BibitemShut
  {NoStop}%
\bibitem [{\citenamefont {Backes}\ \emph {et~al.}(2021)\citenamefont {Backes}
  \emph {et~al.}}]{HAYSTAC:2020kwv}%
  \BibitemOpen
  \bibfield  {author} {\bibinfo {author} {\bibfnamefont {K.~M.}\ \bibnamefont
  {Backes}} \emph {et~al.} (\bibinfo {collaboration} {HAYSTAC}),\ }\bibfield
  {title} {\enquote {\bibinfo {title} {{A quantum-enhanced search for dark
  matter axions}},}\ }\href {\doibase 10.1038/s41586-021-03226-7} {\bibfield
  {journal} {\bibinfo  {journal} {Nature}\ }\textbf {\bibinfo {volume} {590}},\
  \bibinfo {pages} {238--242} (\bibinfo {year} {2021})},\ \Eprint
  {http://arxiv.org/abs/2008.01853} {arXiv:2008.01853 [quant-ph]} \BibitemShut
  {NoStop}%
\bibitem [{\citenamefont {Kwon}\ \emph {et~al.}(2021)\citenamefont {Kwon} \emph
  {et~al.}}]{CAPP:2020utb}%
  \BibitemOpen
  \bibfield  {author} {\bibinfo {author} {\bibfnamefont {Ohjoon}\ \bibnamefont
  {Kwon}} \emph {et~al.} (\bibinfo {collaboration} {CAPP}),\ }\bibfield
  {title} {\enquote {\bibinfo {title} {{First Results from an Axion Haloscope
  at CAPP around 10.7 $\mu$eV}},}\ }\href {\doibase
  10.1103/PhysRevLett.126.191802} {\bibfield  {journal} {\bibinfo  {journal}
  {Phys. Rev. Lett.}\ }\textbf {\bibinfo {volume} {126}},\ \bibinfo {pages}
  {191802} (\bibinfo {year} {2021})},\ \Eprint
  {http://arxiv.org/abs/2012.10764} {arXiv:2012.10764 [hep-ex]} \BibitemShut
  {NoStop}%
\bibitem [{\citenamefont {Cervantes}\ \emph
  {et~al.}(2022{\natexlab{a}})\citenamefont {Cervantes} \emph
  {et~al.}}]{Cervantes:2022epl}%
  \BibitemOpen
  \bibfield  {author} {\bibinfo {author} {\bibfnamefont {R.}~\bibnamefont
  {Cervantes}} \emph {et~al.},\ }\bibfield  {title} {\enquote {\bibinfo {title}
  {{ADMX-Orpheus first search for 70\,\,\ensuremath{\mu}eV dark photon dark
  matter: Detailed design, operations, and analysis}},}\ }\href {\doibase
  10.1103/PhysRevD.106.102002} {\bibfield  {journal} {\bibinfo  {journal}
  {Phys. Rev. D}\ }\textbf {\bibinfo {volume} {106}},\ \bibinfo {pages}
  {102002} (\bibinfo {year} {2022}{\natexlab{a}})},\ \Eprint
  {http://arxiv.org/abs/2204.09475} {arXiv:2204.09475 [hep-ex]} \BibitemShut
  {NoStop}%
\bibitem [{\citenamefont {Cervantes}\ \emph {et~al.}(2024)\citenamefont
  {Cervantes} \emph {et~al.}}]{Cervantes:2022gtv}%
  \BibitemOpen
  \bibfield  {author} {\bibinfo {author} {\bibfnamefont {Raphael}\ \bibnamefont
  {Cervantes}} \emph {et~al.},\ }\bibfield  {title} {\enquote {\bibinfo {title}
  {{Deepest sensitivity to wavelike dark photon dark matter with
  superconducting radio frequency cavities}},}\ }\href {\doibase
  10.1103/PhysRevD.110.043022} {\bibfield  {journal} {\bibinfo  {journal}
  {Phys. Rev. D}\ }\textbf {\bibinfo {volume} {110}},\ \bibinfo {pages}
  {043022} (\bibinfo {year} {2024})},\ \Eprint
  {http://arxiv.org/abs/2208.03183} {arXiv:2208.03183 [hep-ex]} \BibitemShut
  {NoStop}%
\bibitem [{\citenamefont {McAllister}\ \emph {et~al.}(2024)\citenamefont
  {McAllister}, \citenamefont {Quiskamp}, \citenamefont {O'Hare}, \citenamefont
  {Altin}, \citenamefont {Ivanov}, \citenamefont {Goryachev},\ and\
  \citenamefont {Tobar}}]{McAllister:2022ibe}%
  \BibitemOpen
  \bibfield  {author} {\bibinfo {author} {\bibfnamefont {Ben~T.}\ \bibnamefont
  {McAllister}}, \bibinfo {author} {\bibfnamefont {Aaron}\ \bibnamefont
  {Quiskamp}}, \bibinfo {author} {\bibfnamefont {Ciaran A.~J.}\ \bibnamefont
  {O'Hare}}, \bibinfo {author} {\bibfnamefont {Paul}\ \bibnamefont {Altin}},
  \bibinfo {author} {\bibfnamefont {Eugene~N.}\ \bibnamefont {Ivanov}},
  \bibinfo {author} {\bibfnamefont {Maxim}\ \bibnamefont {Goryachev}}, \ and\
  \bibinfo {author} {\bibfnamefont {Michael~E.}\ \bibnamefont {Tobar}},\
  }\bibfield  {title} {\enquote {\bibinfo {title} {{Limits on Dark Photons,
  Scalars, and Axion-Electromagnetodynamics with the ORGAN Experiment}},}\
  }\href {\doibase 10.1002/andp.202200622} {\bibfield  {journal} {\bibinfo
  {journal} {Annalen Phys.}\ }\textbf {\bibinfo {volume} {536}},\ \bibinfo
  {pages} {2200622} (\bibinfo {year} {2024})},\ \Eprint
  {http://arxiv.org/abs/2212.01971} {arXiv:2212.01971 [hep-ph]} \BibitemShut
  {NoStop}%
\bibitem [{\citenamefont {Chang}\ \emph {et~al.}(2022)\citenamefont {Chang}
  \emph {et~al.}}]{TASEH:2022vvu}%
  \BibitemOpen
  \bibfield  {author} {\bibinfo {author} {\bibfnamefont {Hsin}\ \bibnamefont
  {Chang}} \emph {et~al.} (\bibinfo {collaboration} {TASEH}),\ }\bibfield
  {title} {\enquote {\bibinfo {title} {{First Results from the Taiwan Axion
  Search Experiment with a Haloscope at 19.6\,\,\ensuremath{\mu}eV}},}\ }\href
  {\doibase 10.1103/PhysRevLett.129.111802} {\bibfield  {journal} {\bibinfo
  {journal} {Phys. Rev. Lett.}\ }\textbf {\bibinfo {volume} {129}},\ \bibinfo
  {pages} {111802} (\bibinfo {year} {2022})},\ \Eprint
  {http://arxiv.org/abs/2205.05574} {arXiv:2205.05574 [hep-ex]} \BibitemShut
  {NoStop}%
\bibitem [{\citenamefont {Cervantes}\ \emph
  {et~al.}(2022{\natexlab{b}})\citenamefont {Cervantes} \emph
  {et~al.}}]{Cervantes:2022yzp}%
  \BibitemOpen
  \bibfield  {author} {\bibinfo {author} {\bibfnamefont {R.}~\bibnamefont
  {Cervantes}} \emph {et~al.},\ }\bibfield  {title} {\enquote {\bibinfo {title}
  {{Search for 70\,\,\ensuremath{\mu}eV Dark Photon Dark Matter with a
  Dielectrically Loaded Multiwavelength Microwave Cavity}},}\ }\href {\doibase
  10.1103/PhysRevLett.129.201301} {\bibfield  {journal} {\bibinfo  {journal}
  {Phys. Rev. Lett.}\ }\textbf {\bibinfo {volume} {129}},\ \bibinfo {pages}
  {201301} (\bibinfo {year} {2022}{\natexlab{b}})},\ \Eprint
  {http://arxiv.org/abs/2204.03818} {arXiv:2204.03818 [hep-ex]} \BibitemShut
  {NoStop}%
\bibitem [{\citenamefont {Schneemann}\ \emph {et~al.}(2023)\citenamefont
  {Schneemann}, \citenamefont {Schmieden},\ and\ \citenamefont
  {Schott}}]{Schneemann:2023bqc}%
  \BibitemOpen
  \bibfield  {author} {\bibinfo {author} {\bibfnamefont {Tim}\ \bibnamefont
  {Schneemann}}, \bibinfo {author} {\bibfnamefont {Kristof}\ \bibnamefont
  {Schmieden}}, \ and\ \bibinfo {author} {\bibfnamefont {Matthias}\
  \bibnamefont {Schott}},\ }\bibfield  {title} {\enquote {\bibinfo {title}
  {{First results of the SUPAX Experiment: Probing Dark Photons}},}\
  }\href@noop {} {\  (\bibinfo {year} {2023})},\ \Eprint
  {http://arxiv.org/abs/2308.08337} {arXiv:2308.08337 [hep-ex]} \BibitemShut
  {NoStop}%
\bibitem [{\citenamefont {Tang}\ \emph {et~al.}(2024)\citenamefont {Tang} \emph
  {et~al.}}]{SHANHE:2023kxz}%
  \BibitemOpen
  \bibfield  {author} {\bibinfo {author} {\bibfnamefont {Zhenxing}\
  \bibnamefont {Tang}} \emph {et~al.} (\bibinfo {collaboration} {SHANHE}),\
  }\bibfield  {title} {\enquote {\bibinfo {title} {{First Scan Search for Dark
  Photon Dark Matter with a Tunable Superconducting Radio-Frequency Cavity}},}\
  }\href {\doibase 10.1103/PhysRevLett.133.021005} {\bibfield  {journal}
  {\bibinfo  {journal} {Phys. Rev. Lett.}\ }\textbf {\bibinfo {volume} {133}},\
  \bibinfo {pages} {021005} (\bibinfo {year} {2024})},\ \Eprint
  {http://arxiv.org/abs/2305.09711} {arXiv:2305.09711 [hep-ex]} \BibitemShut
  {NoStop}%
\bibitem [{\citenamefont {Ahn}\ \emph {et~al.}(2024)\citenamefont {Ahn} \emph
  {et~al.}}]{CAPP:2024dtx}%
  \BibitemOpen
  \bibfield  {author} {\bibinfo {author} {\bibfnamefont {Saebyeok}\
  \bibnamefont {Ahn}} \emph {et~al.} (\bibinfo {collaboration} {CAPP}),\
  }\bibfield  {title} {\enquote {\bibinfo {title} {{Extensive Search for Axion
  Dark Matter over 1~GHz with CAPP\textquoteright{}S Main Axion Experiment}},}\
  }\href {\doibase 10.1103/PhysRevX.14.031023} {\bibfield  {journal} {\bibinfo
  {journal} {Phys. Rev. X}\ }\textbf {\bibinfo {volume} {14}},\ \bibinfo
  {pages} {031023} (\bibinfo {year} {2024})},\ \Eprint
  {http://arxiv.org/abs/2402.12892} {arXiv:2402.12892 [hep-ex]} \BibitemShut
  {NoStop}%
\bibitem [{\citenamefont {He}\ \emph {et~al.}(2024{\natexlab{a}})\citenamefont
  {He} \emph {et~al.}}]{He:2024fzj}%
  \BibitemOpen
  \bibfield  {author} {\bibinfo {author} {\bibfnamefont {Dong}\ \bibnamefont
  {He}} \emph {et~al.},\ }\bibfield  {title} {\enquote {\bibinfo {title}
  {{Calibration of the cryogenic measurement system of a resonant haloscope
  cavity*}},}\ }\href {\doibase 10.1088/1674-1137/ad4267} {\bibfield  {journal}
  {\bibinfo  {journal} {Chin. Phys. C}\ }\textbf {\bibinfo {volume} {48}},\
  \bibinfo {pages} {073004} (\bibinfo {year} {2024}{\natexlab{a}})},\ \Eprint
  {http://arxiv.org/abs/2404.10264} {arXiv:2404.10264 [hep-ex]} \BibitemShut
  {NoStop}%
\bibitem [{\citenamefont {He}\ \emph {et~al.}(2024{\natexlab{b}})\citenamefont
  {He} \emph {et~al.}}]{APEX:2024jxw}%
  \BibitemOpen
  \bibfield  {author} {\bibinfo {author} {\bibfnamefont {Dong}\ \bibnamefont
  {He}} \emph {et~al.} (\bibinfo {collaboration} {APEX}),\ }\bibfield  {title}
  {\enquote {\bibinfo {title} {{Dark photon constraints from a 7.139~GHz cavity
  haloscope experiment}},}\ }\href {\doibase 10.1103/PhysRevD.110.L021101}
  {\bibfield  {journal} {\bibinfo  {journal} {Phys. Rev. D}\ }\textbf {\bibinfo
  {volume} {110}},\ \bibinfo {pages} {L021101} (\bibinfo {year}
  {2024}{\natexlab{b}})},\ \Eprint {http://arxiv.org/abs/2404.00908}
  {arXiv:2404.00908 [hep-ex]} \BibitemShut {NoStop}%
\bibitem [{\citenamefont {Purcell}(1946)}]{Purcell:1946}%
  \BibitemOpen
  \bibfield  {author} {\bibinfo {author} {\bibfnamefont {E.~M.}\ \bibnamefont
  {Purcell}},\ }\bibfield  {title} {\enquote {\bibinfo {title} {Spontaneous
  emission probabilities at radio frequencies},}\ }\href@noop {} {\bibfield
  {journal} {\bibinfo  {journal} {Phys. Rev.}\ }\textbf {\bibinfo {volume}
  {69}},\ \bibinfo {pages} {681} (\bibinfo {year} {1946})}\BibitemShut
  {NoStop}%
\bibitem [{\citenamefont {Tkachev}(1987)}]{Tkachev:1987cd}%
  \BibitemOpen
  \bibfield  {author} {\bibinfo {author} {\bibfnamefont {I.~I.}\ \bibnamefont
  {Tkachev}},\ }\bibfield  {title} {\enquote {\bibinfo {title} {{An Axionic
  Laser in the Center of a Galaxy?}}}\ }\href {\doibase
  10.1016/0370-2693(87)91318-9} {\bibfield  {journal} {\bibinfo  {journal}
  {Phys. Lett. B}\ }\textbf {\bibinfo {volume} {191}},\ \bibinfo {pages}
  {41--45} (\bibinfo {year} {1987})}\BibitemShut {NoStop}%
\bibitem [{\citenamefont {Kephart}\ and\ \citenamefont
  {Weiler}(1995)}]{Kephart:1994uy}%
  \BibitemOpen
  \bibfield  {author} {\bibinfo {author} {\bibfnamefont {Thomas~W.}\
  \bibnamefont {Kephart}}\ and\ \bibinfo {author} {\bibfnamefont {Thomas~J.}\
  \bibnamefont {Weiler}},\ }\bibfield  {title} {\enquote {\bibinfo {title}
  {{Stimulated radiation from axion cluster evolution}},}\ }\href {\doibase
  10.1103/PhysRevD.52.3226} {\bibfield  {journal} {\bibinfo  {journal} {Phys.
  Rev. D}\ }\textbf {\bibinfo {volume} {52}},\ \bibinfo {pages} {3226--3238}
  (\bibinfo {year} {1995})}\BibitemShut {NoStop}%
\bibitem [{\citenamefont {Rosa}\ and\ \citenamefont
  {Kephart}(2018)}]{Rosa:2017ury}%
  \BibitemOpen
  \bibfield  {author} {\bibinfo {author} {\bibfnamefont {Jo\~ao~G.}\
  \bibnamefont {Rosa}}\ and\ \bibinfo {author} {\bibfnamefont {Thomas~W.}\
  \bibnamefont {Kephart}},\ }\bibfield  {title} {\enquote {\bibinfo {title}
  {{Stimulated Axion Decay in Superradiant Clouds around Primordial Black
  Holes}},}\ }\href {\doibase 10.1103/PhysRevLett.120.231102} {\bibfield
  {journal} {\bibinfo  {journal} {Phys. Rev. Lett.}\ }\textbf {\bibinfo
  {volume} {120}},\ \bibinfo {pages} {231102} (\bibinfo {year} {2018})},\
  \Eprint {http://arxiv.org/abs/1709.06581} {arXiv:1709.06581 [gr-qc]}
  \BibitemShut {NoStop}%
\bibitem [{\citenamefont {Caputo}\ \emph {et~al.}(2019)\citenamefont {Caputo},
  \citenamefont {Regis}, \citenamefont {Taoso},\ and\ \citenamefont
  {Witte}}]{Caputo:2018vmy}%
  \BibitemOpen
  \bibfield  {author} {\bibinfo {author} {\bibfnamefont {Andrea}\ \bibnamefont
  {Caputo}}, \bibinfo {author} {\bibfnamefont {Marco}\ \bibnamefont {Regis}},
  \bibinfo {author} {\bibfnamefont {Marco}\ \bibnamefont {Taoso}}, \ and\
  \bibinfo {author} {\bibfnamefont {Samuel~J.}\ \bibnamefont {Witte}},\
  }\bibfield  {title} {\enquote {\bibinfo {title} {{Detecting the Stimulated
  Decay of Axions at RadioFrequencies}},}\ }\href {\doibase
  10.1088/1475-7516/2019/03/027} {\bibfield  {journal} {\bibinfo  {journal}
  {JCAP}\ }\textbf {\bibinfo {volume} {03}},\ \bibinfo {pages} {027} (\bibinfo
  {year} {2019})},\ \Eprint {http://arxiv.org/abs/1811.08436} {arXiv:1811.08436
  [hep-ph]} \BibitemShut {NoStop}%
\bibitem [{\citenamefont {Arza}(2019)}]{Arza:2018dcy}%
  \BibitemOpen
  \bibfield  {author} {\bibinfo {author} {\bibfnamefont {Ariel}\ \bibnamefont
  {Arza}},\ }\bibfield  {title} {\enquote {\bibinfo {title} {{Photon
  enhancement in a homogeneous axion dark matter background}},}\ }\href
  {\doibase 10.1140/epjc/s10052-019-6759-7} {\bibfield  {journal} {\bibinfo
  {journal} {Eur. Phys. J. C}\ }\textbf {\bibinfo {volume} {79}},\ \bibinfo
  {pages} {250} (\bibinfo {year} {2019})},\ \Eprint
  {http://arxiv.org/abs/1810.03722} {arXiv:1810.03722 [hep-ph]} \BibitemShut
  {NoStop}%
\bibitem [{\citenamefont {Arza}\ and\ \citenamefont
  {Sikivie}(2019)}]{Arza:2019nta}%
  \BibitemOpen
  \bibfield  {author} {\bibinfo {author} {\bibfnamefont {Ariel}\ \bibnamefont
  {Arza}}\ and\ \bibinfo {author} {\bibfnamefont {Pierre}\ \bibnamefont
  {Sikivie}},\ }\bibfield  {title} {\enquote {\bibinfo {title} {{Production and
  detection of an axion dark matter echo}},}\ }\href {\doibase
  10.1103/PhysRevLett.123.131804} {\bibfield  {journal} {\bibinfo  {journal}
  {Phys. Rev. Lett.}\ }\textbf {\bibinfo {volume} {123}},\ \bibinfo {pages}
  {131804} (\bibinfo {year} {2019})},\ \Eprint
  {http://arxiv.org/abs/1902.00114} {arXiv:1902.00114 [hep-ph]} \BibitemShut
  {NoStop}%
\bibitem [{\citenamefont {Arza}\ and\ \citenamefont
  {Todarello}(2022)}]{Arza:2021nec}%
  \BibitemOpen
  \bibfield  {author} {\bibinfo {author} {\bibfnamefont {Ariel}\ \bibnamefont
  {Arza}}\ and\ \bibinfo {author} {\bibfnamefont {Elisa}\ \bibnamefont
  {Todarello}},\ }\bibfield  {title} {\enquote {\bibinfo {title} {{Axion dark
  matter echo: A detailed analysis}},}\ }\href {\doibase
  10.1103/PhysRevD.105.023023} {\bibfield  {journal} {\bibinfo  {journal}
  {Phys. Rev. D}\ }\textbf {\bibinfo {volume} {105}},\ \bibinfo {pages}
  {023023} (\bibinfo {year} {2022})},\ \Eprint
  {http://arxiv.org/abs/2108.00195} {arXiv:2108.00195 [hep-ph]} \BibitemShut
  {NoStop}%
\bibitem [{\citenamefont {Arza}\ and\ \citenamefont
  {Todarello}(2021)}]{Arza:2021zqc}%
  \BibitemOpen
  \bibfield  {author} {\bibinfo {author} {\bibfnamefont {Ariel}\ \bibnamefont
  {Arza}}\ and\ \bibinfo {author} {\bibfnamefont {Elisa}\ \bibnamefont
  {Todarello}},\ }\bibfield  {title} {\enquote {\bibinfo {title} {{The Echo
  Method for Axion Dark Matter Detection}},}\ }\href {\doibase
  10.3390/sym13112150} {\bibfield  {journal} {\bibinfo  {journal} {Symmetry}\
  }\textbf {\bibinfo {volume} {13}},\ \bibinfo {pages} {2150} (\bibinfo {year}
  {2021})}\BibitemShut {NoStop}%
\bibitem [{\citenamefont {Arza}\ \emph {et~al.}(2023)\citenamefont {Arza},
  \citenamefont {Kryemadhi},\ and\ \citenamefont {Zioutas}}]{Arza:2022dng}%
  \BibitemOpen
  \bibfield  {author} {\bibinfo {author} {\bibfnamefont {Ariel}\ \bibnamefont
  {Arza}}, \bibinfo {author} {\bibfnamefont {Abaz}\ \bibnamefont {Kryemadhi}},
  \ and\ \bibinfo {author} {\bibfnamefont {Konstantin}\ \bibnamefont
  {Zioutas}},\ }\bibfield  {title} {\enquote {\bibinfo {title} {{Searching for
  axion streams with the echo method}},}\ }\href {\doibase
  10.1103/PhysRevD.108.083001} {\bibfield  {journal} {\bibinfo  {journal}
  {Phys. Rev. D}\ }\textbf {\bibinfo {volume} {108}},\ \bibinfo {pages}
  {083001} (\bibinfo {year} {2023})},\ \Eprint
  {http://arxiv.org/abs/2212.10905} {arXiv:2212.10905 [hep-ph]} \BibitemShut
  {NoStop}%
\bibitem [{\citenamefont {Arza}\ \emph {et~al.}(2024)\citenamefont {Arza},
  \citenamefont {Guo}, \citenamefont {Wu}, \citenamefont {Yang}, \citenamefont
  {Yang}, \citenamefont {Yuan},\ and\ \citenamefont {Zhu}}]{Arza:2023rcs}%
  \BibitemOpen
  \bibfield  {author} {\bibinfo {author} {\bibfnamefont {Ariel}\ \bibnamefont
  {Arza}}, \bibinfo {author} {\bibfnamefont {Quan}\ \bibnamefont {Guo}},
  \bibinfo {author} {\bibfnamefont {Lei}\ \bibnamefont {Wu}}, \bibinfo {author}
  {\bibfnamefont {Qiaoli}\ \bibnamefont {Yang}}, \bibinfo {author}
  {\bibfnamefont {Xiaolong}\ \bibnamefont {Yang}}, \bibinfo {author}
  {\bibfnamefont {Qiang}\ \bibnamefont {Yuan}}, \ and\ \bibinfo {author}
  {\bibfnamefont {Bin}\ \bibnamefont {Zhu}},\ }\bibfield  {title} {\enquote
  {\bibinfo {title} {{Listening for echo from the stimulated axion decay with
  the 21 centimeter array}},}\ }\href {\doibase 10.1016/j.scib.2024.08.003}
  {\bibfield  {journal} {\bibinfo  {journal} {Sci. Bull.}\ }\textbf {\bibinfo
  {volume} {69}},\ \bibinfo {pages} {2971--2973} (\bibinfo {year} {2024})},\
  \Eprint {http://arxiv.org/abs/2309.06857} {arXiv:2309.06857 [hep-ph]}
  \BibitemShut {NoStop}%
\bibitem [{\citenamefont {Berlin}\ \emph {et~al.}(2020)\citenamefont {Berlin},
  \citenamefont {D'Agnolo}, \citenamefont {Ellis}, \citenamefont {Nantista},
  \citenamefont {Neilson}, \citenamefont {Schuster}, \citenamefont {Tantawi},
  \citenamefont {Toro},\ and\ \citenamefont {Zhou}}]{Berlin:2019ahk}%
  \BibitemOpen
  \bibfield  {author} {\bibinfo {author} {\bibfnamefont {Asher}\ \bibnamefont
  {Berlin}}, \bibinfo {author} {\bibfnamefont {Raffaele~Tito}\ \bibnamefont
  {D'Agnolo}}, \bibinfo {author} {\bibfnamefont {Sebastian A.~R.}\ \bibnamefont
  {Ellis}}, \bibinfo {author} {\bibfnamefont {Christopher}\ \bibnamefont
  {Nantista}}, \bibinfo {author} {\bibfnamefont {Jeffrey}\ \bibnamefont
  {Neilson}}, \bibinfo {author} {\bibfnamefont {Philip}\ \bibnamefont
  {Schuster}}, \bibinfo {author} {\bibfnamefont {Sami}\ \bibnamefont
  {Tantawi}}, \bibinfo {author} {\bibfnamefont {Natalia}\ \bibnamefont {Toro}},
  \ and\ \bibinfo {author} {\bibfnamefont {Kevin}\ \bibnamefont {Zhou}},\
  }\bibfield  {title} {\enquote {\bibinfo {title} {{Axion Dark Matter Detection
  by Superconducting Resonant Frequency Conversion}},}\ }\href {\doibase
  10.1007/JHEP07(2020)088} {\bibfield  {journal} {\bibinfo  {journal} {JHEP}\
  }\textbf {\bibinfo {volume} {07}},\ \bibinfo {pages} {088} (\bibinfo {year}
  {2020})},\ \Eprint {http://arxiv.org/abs/1912.11048} {arXiv:1912.11048
  [hep-ph]} \BibitemShut {NoStop}%
\bibitem [{\citenamefont {Lasenby}(2020)}]{Lasenby:2019prg}%
  \BibitemOpen
  \bibfield  {author} {\bibinfo {author} {\bibfnamefont {Robert}\ \bibnamefont
  {Lasenby}},\ }\bibfield  {title} {\enquote {\bibinfo {title} {{Microwave
  cavity searches for low-frequency axion dark matter}},}\ }\href {\doibase
  10.1103/PhysRevD.102.015008} {\bibfield  {journal} {\bibinfo  {journal}
  {Phys. Rev. D}\ }\textbf {\bibinfo {volume} {102}},\ \bibinfo {pages}
  {015008} (\bibinfo {year} {2020})},\ \Eprint
  {http://arxiv.org/abs/1912.11056} {arXiv:1912.11056 [hep-ph]} \BibitemShut
  {NoStop}%
\bibitem [{\citenamefont {Berlin}\ \emph {et~al.}(2021)\citenamefont {Berlin},
  \citenamefont {D'Agnolo}, \citenamefont {Ellis},\ and\ \citenamefont
  {Zhou}}]{Berlin:2020vrk}%
  \BibitemOpen
  \bibfield  {author} {\bibinfo {author} {\bibfnamefont {Asher}\ \bibnamefont
  {Berlin}}, \bibinfo {author} {\bibfnamefont {Raffaele~Tito}\ \bibnamefont
  {D'Agnolo}}, \bibinfo {author} {\bibfnamefont {Sebastian A.~R.}\ \bibnamefont
  {Ellis}}, \ and\ \bibinfo {author} {\bibfnamefont {Kevin}\ \bibnamefont
  {Zhou}},\ }\bibfield  {title} {\enquote {\bibinfo {title} {{Heterodyne
  broadband detection of axion dark matter}},}\ }\href {\doibase
  10.1103/PhysRevD.104.L111701} {\bibfield  {journal} {\bibinfo  {journal}
  {Phys. Rev. D}\ }\textbf {\bibinfo {volume} {104}},\ \bibinfo {pages}
  {L111701} (\bibinfo {year} {2021})},\ \Eprint
  {http://arxiv.org/abs/2007.15656} {arXiv:2007.15656 [hep-ph]} \BibitemShut
  {NoStop}%
\bibitem [{\citenamefont {Li}\ \emph {et~al.}(2025)\citenamefont {Li},
  \citenamefont {Zhou}, \citenamefont {Oriunno}, \citenamefont {Berlin},
  \citenamefont {Calatroni}, \citenamefont {D'Agnolo}, \citenamefont {Ellis},
  \citenamefont {Schuster}, \citenamefont {Tantawi},\ and\ \citenamefont
  {Toro}}]{Li:2025pyi}%
  \BibitemOpen
  \bibfield  {author} {\bibinfo {author} {\bibfnamefont {Zenghai}\ \bibnamefont
  {Li}}, \bibinfo {author} {\bibfnamefont {Kevin}\ \bibnamefont {Zhou}},
  \bibinfo {author} {\bibfnamefont {Marco}\ \bibnamefont {Oriunno}}, \bibinfo
  {author} {\bibfnamefont {Asher}\ \bibnamefont {Berlin}}, \bibinfo {author}
  {\bibfnamefont {Sergio}\ \bibnamefont {Calatroni}}, \bibinfo {author}
  {\bibfnamefont {Raffaele~Tito}\ \bibnamefont {D'Agnolo}}, \bibinfo {author}
  {\bibfnamefont {Sebastian A.~R.}\ \bibnamefont {Ellis}}, \bibinfo {author}
  {\bibfnamefont {Philip}\ \bibnamefont {Schuster}}, \bibinfo {author}
  {\bibfnamefont {Sami~G.}\ \bibnamefont {Tantawi}}, \ and\ \bibinfo {author}
  {\bibfnamefont {Natalia}\ \bibnamefont {Toro}},\ }\bibfield  {title}
  {\enquote {\bibinfo {title} {{A Prototype Hybrid Mode Cavity for Heterodyne
  Axion Detection}},}\ }\href@noop {} {\  (\bibinfo {year} {2025})},\ \Eprint
  {http://arxiv.org/abs/2507.07173} {arXiv:2507.07173 [physics.ins-det]}
  \BibitemShut {NoStop}%
\bibitem [{\citenamefont {Sikivie}(2010)}]{Sikivie:2010fa}%
  \BibitemOpen
  \bibfield  {author} {\bibinfo {author} {\bibfnamefont {P.}~\bibnamefont
  {Sikivie}},\ }\bibfield  {title} {\enquote {\bibinfo {title}
  {{Superconducting Radio Frequency Cavities as Axion Dark Matter
  Detectors}},}\ }\href@noop {} {\  (\bibinfo {year} {2010})},\ \Eprint
  {http://arxiv.org/abs/1009.0762} {arXiv:1009.0762 [hep-ph]} \BibitemShut
  {NoStop}%
\bibitem [{\citenamefont {Janish}\ \emph {et~al.}(2019)\citenamefont {Janish},
  \citenamefont {Narayan}, \citenamefont {Rajendran},\ and\ \citenamefont
  {Riggins}}]{Janish:2019dpr}%
  \BibitemOpen
  \bibfield  {author} {\bibinfo {author} {\bibfnamefont {Ryan}\ \bibnamefont
  {Janish}}, \bibinfo {author} {\bibfnamefont {Vijay}\ \bibnamefont {Narayan}},
  \bibinfo {author} {\bibfnamefont {Surjeet}\ \bibnamefont {Rajendran}}, \ and\
  \bibinfo {author} {\bibfnamefont {Paul}\ \bibnamefont {Riggins}},\ }\bibfield
   {title} {\enquote {\bibinfo {title} {{Axion production and detection with
  superconducting RF cavities}},}\ }\href {\doibase
  10.1103/PhysRevD.100.015036} {\bibfield  {journal} {\bibinfo  {journal}
  {Phys. Rev. D}\ }\textbf {\bibinfo {volume} {100}},\ \bibinfo {pages}
  {015036} (\bibinfo {year} {2019})},\ \Eprint
  {http://arxiv.org/abs/1904.07245} {arXiv:1904.07245 [hep-ph]} \BibitemShut
  {NoStop}%
\bibitem [{\citenamefont {Bogorad}\ \emph {et~al.}(2019)\citenamefont
  {Bogorad}, \citenamefont {Hook}, \citenamefont {Kahn},\ and\ \citenamefont
  {Soreq}}]{Bogorad:2019pbu}%
  \BibitemOpen
  \bibfield  {author} {\bibinfo {author} {\bibfnamefont {Zachary}\ \bibnamefont
  {Bogorad}}, \bibinfo {author} {\bibfnamefont {Anson}\ \bibnamefont {Hook}},
  \bibinfo {author} {\bibfnamefont {Yonatan}\ \bibnamefont {Kahn}}, \ and\
  \bibinfo {author} {\bibfnamefont {Yotam}\ \bibnamefont {Soreq}},\ }\bibfield
  {title} {\enquote {\bibinfo {title} {{Probing Axionlike Particles and the
  Axiverse with Superconducting Radio-Frequency Cavities}},}\ }\href {\doibase
  10.1103/PhysRevLett.123.021801} {\bibfield  {journal} {\bibinfo  {journal}
  {Phys. Rev. Lett.}\ }\textbf {\bibinfo {volume} {123}},\ \bibinfo {pages}
  {021801} (\bibinfo {year} {2019})},\ \Eprint
  {http://arxiv.org/abs/1902.01418} {arXiv:1902.01418 [hep-ph]} \BibitemShut
  {NoStop}%
\bibitem [{\citenamefont {Gao}\ and\ \citenamefont
  {Harnik}(2021)}]{Gao:2020anb}%
  \BibitemOpen
  \bibfield  {author} {\bibinfo {author} {\bibfnamefont {Christina}\
  \bibnamefont {Gao}}\ and\ \bibinfo {author} {\bibfnamefont {Roni}\
  \bibnamefont {Harnik}},\ }\bibfield  {title} {\enquote {\bibinfo {title}
  {{Axion searches with two superconducting radio-frequency cavities}},}\
  }\href {\doibase 10.1007/JHEP07(2021)053} {\bibfield  {journal} {\bibinfo
  {journal} {JHEP}\ }\textbf {\bibinfo {volume} {07}},\ \bibinfo {pages} {053}
  (\bibinfo {year} {2021})},\ \Eprint {http://arxiv.org/abs/2011.01350}
  {arXiv:2011.01350 [hep-ph]} \BibitemShut {NoStop}%
\bibitem [{\citenamefont {Salnikov}\ \emph {et~al.}(2021)\citenamefont
  {Salnikov}, \citenamefont {Satunin}, \citenamefont {Kirpichnikov},\ and\
  \citenamefont {Fitkevich}}]{Salnikov:2020urr}%
  \BibitemOpen
  \bibfield  {author} {\bibinfo {author} {\bibfnamefont {Dmitry}\ \bibnamefont
  {Salnikov}}, \bibinfo {author} {\bibfnamefont {Petr}\ \bibnamefont
  {Satunin}}, \bibinfo {author} {\bibfnamefont {D.~V.}\ \bibnamefont
  {Kirpichnikov}}, \ and\ \bibinfo {author} {\bibfnamefont {Maxim}\
  \bibnamefont {Fitkevich}},\ }\bibfield  {title} {\enquote {\bibinfo {title}
  {{Examining axion-like particles with superconducting radio-frequency
  cavity}},}\ }\href {\doibase 10.1007/jhep03(2021)143} {\bibfield  {journal}
  {\bibinfo  {journal} {JHEP}\ }\textbf {\bibinfo {volume} {03}},\ \bibinfo
  {pages} {143} (\bibinfo {year} {2021})},\ \Eprint
  {http://arxiv.org/abs/2011.12871} {arXiv:2011.12871 [hep-ph]} \BibitemShut
  {NoStop}%
\bibitem [{\citenamefont {Goryachev}\ \emph {et~al.}()\citenamefont
  {Goryachev}, \citenamefont {Mcallister},\ and\ \citenamefont
  {Tobar}}]{Goryachev:2018vjt}%
  \BibitemOpen
  \bibfield  {author} {\bibinfo {author} {\bibfnamefont {Maxim}\ \bibnamefont
  {Goryachev}}, \bibinfo {author} {\bibfnamefont {Ben}\ \bibnamefont
  {Mcallister}}, \ and\ \bibinfo {author} {\bibfnamefont {Michael~E.}\
  \bibnamefont {Tobar}},\ }\bibfield  {title} {\enquote {\bibinfo {title}
  {{Axion detection with precision frequency metrology}},}\ }\href {\doibase
  https://doi.org/10.1016/j.dark.2019.100345} {\
  https://doi.org/10.1016/j.dark.2019.100345},\ \Eprint
  {http://arxiv.org/abs/1806.07141} {arXiv:1806.07141 [physics.ins-det]}
  \BibitemShut {NoStop}%
\bibitem [{\citenamefont {Thomson}\ \emph {et~al.}(2021)\citenamefont
  {Thomson}, \citenamefont {McAllister}, \citenamefont {Goryachev},
  \citenamefont {Ivanov},\ and\ \citenamefont {Tobar}}]{Thomson:2021zvq}%
  \BibitemOpen
  \bibfield  {author} {\bibinfo {author} {\bibfnamefont {Catriona~A.}\
  \bibnamefont {Thomson}}, \bibinfo {author} {\bibfnamefont {Ben~T.}\
  \bibnamefont {McAllister}}, \bibinfo {author} {\bibfnamefont {Maxim}\
  \bibnamefont {Goryachev}}, \bibinfo {author} {\bibfnamefont {Eugene~N.}\
  \bibnamefont {Ivanov}}, \ and\ \bibinfo {author} {\bibfnamefont {Michael~E.}\
  \bibnamefont {Tobar}},\ }\bibfield  {title} {\enquote {\bibinfo {title}
  {{Upconversion Loop Oscillator Axion Detection Experiment: A Precision
  Frequency Interferometric Axion Dark Matter Search with a Cylindrical
  Microwave Cavity}},}\ }\href {\doibase 10.1103/PhysRevLett.127.019901}
  {\bibfield  {journal} {\bibinfo  {journal} {Phys. Rev. Lett.}\ }\textbf
  {\bibinfo {volume} {126}},\ \bibinfo {pages} {081803} (\bibinfo {year}
  {2021})},\ \bibinfo {note} {[Erratum: Phys.Rev.Lett. 127, 019901 (2021)]},\
  \Eprint {http://arxiv.org/abs/1912.07751} {arXiv:1912.07751 [hep-ex]}
  \BibitemShut {NoStop}%
\bibitem [{\citenamefont {Giaccone}\ \emph {et~al.}(2022)\citenamefont
  {Giaccone} \emph {et~al.}}]{Giaccone:2022pke}%
  \BibitemOpen
  \bibfield  {author} {\bibinfo {author} {\bibfnamefont {B.}~\bibnamefont
  {Giaccone}} \emph {et~al.},\ }\bibfield  {title} {\enquote {\bibinfo {title}
  {{Design of axion and axion dark matter searches based on ultra high Q SRF
  cavities}},}\ }\href@noop {} {\  (\bibinfo {year} {2022})},\ \Eprint
  {http://arxiv.org/abs/2207.11346} {arXiv:2207.11346 [hep-ex]} \BibitemShut
  {NoStop}%
\bibitem [{Sup()}]{SupplementalMaterial}%
  \BibitemOpen
  \href@noop {} {\ }\bibinfo {note} {See Supplemental Material below for
  detailed calculation of the signal power}\BibitemShut {NoStop}%
\bibitem [{\citenamefont {Gardiner}\ and\ \citenamefont
  {Collett}(1985)}]{PhysRevA.31.3761}%
  \BibitemOpen
  \bibfield  {author} {\bibinfo {author} {\bibfnamefont {C.~W.}\ \bibnamefont
  {Gardiner}}\ and\ \bibinfo {author} {\bibfnamefont {M.~J.}\ \bibnamefont
  {Collett}},\ }\bibfield  {title} {\enquote {\bibinfo {title} {Input and
  output in damped quantum systems: Quantum stochastic differential equations
  and the master equation},}\ }\href {\doibase 10.1103/PhysRevA.31.3761}
  {\bibfield  {journal} {\bibinfo  {journal} {Phys. Rev. A}\ }\textbf {\bibinfo
  {volume} {31}},\ \bibinfo {pages} {3761--3774} (\bibinfo {year}
  {1985})}\BibitemShut {NoStop}%
\bibitem [{\citenamefont {Chaudhuri}\ \emph {et~al.}(2018)\citenamefont
  {Chaudhuri}, \citenamefont {Irwin}, \citenamefont {Graham},\ and\
  \citenamefont {Mardon}}]{Chaudhuri:2018rqn}%
  \BibitemOpen
  \bibfield  {author} {\bibinfo {author} {\bibfnamefont {Saptarshi}\
  \bibnamefont {Chaudhuri}}, \bibinfo {author} {\bibfnamefont {Kent}\
  \bibnamefont {Irwin}}, \bibinfo {author} {\bibfnamefont {Peter~W.}\
  \bibnamefont {Graham}}, \ and\ \bibinfo {author} {\bibfnamefont {Jeremy}\
  \bibnamefont {Mardon}},\ }\bibfield  {title} {\enquote {\bibinfo {title}
  {{Optimal Impedance Matching and Quantum Limits of Electromagnetic Axion and
  Hidden-Photon Dark Matter Searches}},}\ }\href@noop {} {\  (\bibinfo {year}
  {2018})},\ \Eprint {http://arxiv.org/abs/1803.01627} {arXiv:1803.01627
  [hep-ph]} \BibitemShut {NoStop}%
\bibitem [{\citenamefont {O'Hare}(2020)}]{AxionLimits}%
  \BibitemOpen
  \bibfield  {author} {\bibinfo {author} {\bibfnamefont {Ciaran}\ \bibnamefont
  {O'Hare}},\ }\href {\doibase 10.5281/zenodo.3932430} {\enquote {\bibinfo
  {title} {cajohare/axionlimits: Axionlimits},}\ }\bibinfo {howpublished}
  {\url{https://cajohare.github.io/AxionLimits/}} (\bibinfo {year}
  {2020})\BibitemShut {NoStop}%
\bibitem [{\citenamefont {Anastassopoulos}\ \emph {et~al.}(2017)\citenamefont
  {Anastassopoulos} \emph {et~al.}}]{CAST:2017uph}%
  \BibitemOpen
  \bibfield  {author} {\bibinfo {author} {\bibfnamefont {V.}~\bibnamefont
  {Anastassopoulos}} \emph {et~al.} (\bibinfo {collaboration} {CAST}),\
  }\bibfield  {title} {\enquote {\bibinfo {title} {{New CAST Limit on the
  Axion-Photon Interaction}},}\ }\href {\doibase 10.1038/nphys4109} {\bibfield
  {journal} {\bibinfo  {journal} {Nature Phys.}\ }\textbf {\bibinfo {volume}
  {13}},\ \bibinfo {pages} {584--590} (\bibinfo {year} {2017})},\ \Eprint
  {http://arxiv.org/abs/1705.02290} {arXiv:1705.02290 [hep-ex]} \BibitemShut
  {NoStop}%
\bibitem [{\citenamefont {Noordhuis}\ \emph {et~al.}(2023)\citenamefont
  {Noordhuis}, \citenamefont {Prabhu}, \citenamefont {Witte}, \citenamefont
  {Chen}, \citenamefont {Cruz},\ and\ \citenamefont
  {Weniger}}]{Noordhuis:2022ljw}%
  \BibitemOpen
  \bibfield  {author} {\bibinfo {author} {\bibfnamefont {Dion}\ \bibnamefont
  {Noordhuis}}, \bibinfo {author} {\bibfnamefont {Anirudh}\ \bibnamefont
  {Prabhu}}, \bibinfo {author} {\bibfnamefont {Samuel~J.}\ \bibnamefont
  {Witte}}, \bibinfo {author} {\bibfnamefont {Alexander~Y.}\ \bibnamefont
  {Chen}}, \bibinfo {author} {\bibfnamefont {F\'abio}\ \bibnamefont {Cruz}}, \
  and\ \bibinfo {author} {\bibfnamefont {Christoph}\ \bibnamefont {Weniger}},\
  }\bibfield  {title} {\enquote {\bibinfo {title} {Novel constraints on axions
  produced in pulsar polar-cap cascades},}\ }\href {\doibase
  10.1103/PhysRevLett.131.111004} {\bibfield  {journal} {\bibinfo  {journal}
  {Phys. Rev. Lett.}\ }\textbf {\bibinfo {volume} {131}},\ \bibinfo {pages}
  {111004} (\bibinfo {year} {2023})},\ \Eprint
  {http://arxiv.org/abs/2209.09917} {2209.09917} \BibitemShut {NoStop}%
\bibitem [{\citenamefont {Hoof}\ and\ \citenamefont
  {Schulz}(2022)}]{Hoof:2022xbe}%
  \BibitemOpen
  \bibfield  {author} {\bibinfo {author} {\bibfnamefont {Sebastian}\
  \bibnamefont {Hoof}}\ and\ \bibinfo {author} {\bibfnamefont {Lena}\
  \bibnamefont {Schulz}},\ }\bibfield  {title} {\enquote {\bibinfo {title}
  {{Updated constraints on axion-like particles from temporal information in
  supernova SN1987A gamma-ray data}},}\ }\href@noop {} {\  (\bibinfo {year}
  {2022})},\ \Eprint {http://arxiv.org/abs/2212.09764} {arXiv:2212.09764
  [hep-ph]} \BibitemShut {NoStop}%
\bibitem [{\citenamefont {Manzari}\ \emph {et~al.}(2024)\citenamefont
  {Manzari}, \citenamefont {Park}, \citenamefont {Safdi},\ and\ \citenamefont
  {Savoray}}]{Manzari:2024jns}%
  \BibitemOpen
  \bibfield  {author} {\bibinfo {author} {\bibfnamefont {Claudio~Andrea}\
  \bibnamefont {Manzari}}, \bibinfo {author} {\bibfnamefont {Yujin}\
  \bibnamefont {Park}}, \bibinfo {author} {\bibfnamefont {Benjamin~R.}\
  \bibnamefont {Safdi}}, \ and\ \bibinfo {author} {\bibfnamefont {Inbar}\
  \bibnamefont {Savoray}},\ }\bibfield  {title} {\enquote {\bibinfo {title}
  {{Supernova Axions Convert to Gamma Rays in Magnetic Fields of Progenitor
  Stars}},}\ }\href {\doibase 10.1103/PhysRevLett.133.211002} {\bibfield
  {journal} {\bibinfo  {journal} {Phys. Rev. Lett.}\ }\textbf {\bibinfo
  {volume} {133}},\ \bibinfo {pages} {211002} (\bibinfo {year} {2024})},\
  \Eprint {http://arxiv.org/abs/2405.19393} {arXiv:2405.19393 [hep-ph]}
  \BibitemShut {NoStop}%
\bibitem [{\citenamefont {Ruz}\ \emph {et~al.}(2024)\citenamefont {Ruz} \emph
  {et~al.}}]{Ruz:2024gkl}%
  \BibitemOpen
  \bibfield  {author} {\bibinfo {author} {\bibfnamefont {J.}~\bibnamefont
  {Ruz}} \emph {et~al.},\ }\bibfield  {title} {\enquote {\bibinfo {title}
  {{NuSTAR as an Axion Helioscope}},}\ }\href@noop {} {\  (\bibinfo {year}
  {2024})},\ \Eprint {http://arxiv.org/abs/2407.03828} {arXiv:2407.03828
  [astro-ph.CO]} \BibitemShut {NoStop}%
\bibitem [{\citenamefont {Hagmann}\ \emph {et~al.}(1990)\citenamefont
  {Hagmann}, \citenamefont {Sikivie}, \citenamefont {Sullivan},\ and\
  \citenamefont {Tanner}}]{Hagmann}%
  \BibitemOpen
  \bibfield  {author} {\bibinfo {author} {\bibfnamefont {C.}~\bibnamefont
  {Hagmann}}, \bibinfo {author} {\bibfnamefont {P.}~\bibnamefont {Sikivie}},
  \bibinfo {author} {\bibfnamefont {N.~S.}\ \bibnamefont {Sullivan}}, \ and\
  \bibinfo {author} {\bibfnamefont {D.~B.}\ \bibnamefont {Tanner}},\ }\bibfield
   {title} {\enquote {\bibinfo {title} {Results from a search for cosmic
  axions},}\ }\href {\doibase 10.1103/PhysRevD.42.1297} {\bibfield  {journal}
  {\bibinfo  {journal} {Phys. Rev. D}\ }\textbf {\bibinfo {volume} {42}},\
  \bibinfo {pages} {1297--1300} (\bibinfo {year} {1990})}\BibitemShut {NoStop}%
\bibitem [{\citenamefont {Hagmann}\ \emph {et~al.}(1996)\citenamefont {Hagmann}
  \emph {et~al.}}]{Hagmann:1996qd}%
  \BibitemOpen
  \bibfield  {author} {\bibinfo {author} {\bibfnamefont {C.}~\bibnamefont
  {Hagmann}} \emph {et~al.},\ }\bibfield  {title} {\enquote {\bibinfo {title}
  {{First results from a second generation galactic axion experiment}},}\
  }\href {\doibase 10.1016/S0920-5632(96)00516-6} {\bibfield  {journal}
  {\bibinfo  {journal} {Nucl. Phys. B Proc. Suppl.}\ }\textbf {\bibinfo
  {volume} {51}},\ \bibinfo {pages} {209--212} (\bibinfo {year} {1996})},\
  \Eprint {http://arxiv.org/abs/astro-ph/9607022} {arXiv:astro-ph/9607022}
  \BibitemShut {NoStop}%
\bibitem [{\citenamefont {DePanfilis}\ \emph {et~al.}(1987)\citenamefont
  {DePanfilis}, \citenamefont {Melissinos}, \citenamefont {Moskowitz},
  \citenamefont {Rogers}, \citenamefont {Semertzidis}, \citenamefont {Wuensch},
  \citenamefont {Halama}, \citenamefont {Prodell}, \citenamefont {Fowler},\
  and\ \citenamefont {Nezrick}}]{DePanfilis}%
  \BibitemOpen
  \bibfield  {author} {\bibinfo {author} {\bibfnamefont {S.}~\bibnamefont
  {DePanfilis}}, \bibinfo {author} {\bibfnamefont {A.~C.}\ \bibnamefont
  {Melissinos}}, \bibinfo {author} {\bibfnamefont {B.~E.}\ \bibnamefont
  {Moskowitz}}, \bibinfo {author} {\bibfnamefont {J.~T.}\ \bibnamefont
  {Rogers}}, \bibinfo {author} {\bibfnamefont {Y.~K.}\ \bibnamefont
  {Semertzidis}}, \bibinfo {author} {\bibfnamefont {W.~U.}\ \bibnamefont
  {Wuensch}}, \bibinfo {author} {\bibfnamefont {H.~J.}\ \bibnamefont {Halama}},
  \bibinfo {author} {\bibfnamefont {A.~G.}\ \bibnamefont {Prodell}}, \bibinfo
  {author} {\bibfnamefont {W.~B.}\ \bibnamefont {Fowler}}, \ and\ \bibinfo
  {author} {\bibfnamefont {F.~A.}\ \bibnamefont {Nezrick}},\ }\bibfield
  {title} {\enquote {\bibinfo {title} {Limits on the abundance and coupling of
  cosmic axions at 4.5$<{m}_{a}<$5.0 \ensuremath{\mu}ev},}\ }\href {\doibase
  10.1103/PhysRevLett.59.839} {\bibfield  {journal} {\bibinfo  {journal} {Phys.
  Rev. Lett.}\ }\textbf {\bibinfo {volume} {59}},\ \bibinfo {pages} {839--842}
  (\bibinfo {year} {1987})}\BibitemShut {NoStop}%
\bibitem [{\citenamefont {Wuensch}\ \emph {et~al.}(1989)\citenamefont
  {Wuensch}, \citenamefont {De~Panfilis-Wuensch}, \citenamefont {Semertzidis},
  \citenamefont {Rogers}, \citenamefont {Melissinos}, \citenamefont {Halama},
  \citenamefont {Moskowitz}, \citenamefont {Prodell}, \citenamefont {Fowler},\
  and\ \citenamefont {Nezrick}}]{Wuensch:1989sa}%
  \BibitemOpen
  \bibfield  {author} {\bibinfo {author} {\bibfnamefont {Walter}\ \bibnamefont
  {Wuensch}}, \bibinfo {author} {\bibfnamefont {S.}~\bibnamefont
  {De~Panfilis-Wuensch}}, \bibinfo {author} {\bibfnamefont {Y.~K.}\
  \bibnamefont {Semertzidis}}, \bibinfo {author} {\bibfnamefont {J.~T.}\
  \bibnamefont {Rogers}}, \bibinfo {author} {\bibfnamefont {A.~C.}\
  \bibnamefont {Melissinos}}, \bibinfo {author} {\bibfnamefont {H.~J.}\
  \bibnamefont {Halama}}, \bibinfo {author} {\bibfnamefont {B.~E.}\
  \bibnamefont {Moskowitz}}, \bibinfo {author} {\bibfnamefont {A.~G.}\
  \bibnamefont {Prodell}}, \bibinfo {author} {\bibfnamefont {W.~B.}\
  \bibnamefont {Fowler}}, \ and\ \bibinfo {author} {\bibfnamefont {F.~A.}\
  \bibnamefont {Nezrick}},\ }\bibfield  {title} {\enquote {\bibinfo {title}
  {{Results of a Laboratory Search for Cosmic Axions and Other Weakly Coupled
  Light Particles}},}\ }\href {\doibase 10.1103/PhysRevD.40.3153} {\bibfield
  {journal} {\bibinfo  {journal} {Phys. Rev. D}\ }\textbf {\bibinfo {volume}
  {40}},\ \bibinfo {pages} {3153} (\bibinfo {year} {1989})}\BibitemShut
  {NoStop}%
\bibitem [{\citenamefont {Braine}\ \emph {et~al.}(2020)\citenamefont {Braine}
  \emph {et~al.}}]{ADMX:2019uok}%
  \BibitemOpen
  \bibfield  {author} {\bibinfo {author} {\bibfnamefont {T.}~\bibnamefont
  {Braine}} \emph {et~al.} (\bibinfo {collaboration} {ADMX}),\ }\bibfield
  {title} {\enquote {\bibinfo {title} {{Extended Search for the Invisible Axion
  with the Axion Dark Matter Experiment}},}\ }\href {\doibase
  10.1103/PhysRevLett.124.101303} {\bibfield  {journal} {\bibinfo  {journal}
  {Phys. Rev. Lett.}\ }\textbf {\bibinfo {volume} {124}},\ \bibinfo {pages}
  {101303} (\bibinfo {year} {2020})},\ \Eprint
  {http://arxiv.org/abs/1910.08638} {arXiv:1910.08638 [hep-ex]} \BibitemShut
  {NoStop}%
\bibitem [{\citenamefont {Carosi}\ \emph {et~al.}(2025)\citenamefont {Carosi}
  \emph {et~al.}}]{ADMX:2025vom}%
  \BibitemOpen
  \bibfield  {author} {\bibinfo {author} {\bibfnamefont {G.}~\bibnamefont
  {Carosi}} \emph {et~al.} (\bibinfo {collaboration} {ADMX}),\ }\bibfield
  {title} {\enquote {\bibinfo {title} {{Search for Axion Dark Matter from 1.1
  to 1.3 GHz with ADMX}},}\ }\href@noop {} {\  (\bibinfo {year} {2025})},\
  \Eprint {http://arxiv.org/abs/2504.07279} {arXiv:2504.07279 [hep-ex]}
  \BibitemShut {NoStop}%
\bibitem [{\citenamefont {Hipp}(2025)}]{ADMX:2024pxg}%
  \BibitemOpen
  \bibfield  {author} {\bibinfo {author} {\bibfnamefont {A.~T.~{\it et al}}\
  \bibnamefont {Hipp}} (\bibinfo {collaboration} {ADMX Collaboration}),\
  }\bibfield  {title} {\enquote {\bibinfo {title} {Search for nonvirialized
  axions with $3.3\text{ }\ensuremath{-}\text{ }4.2\text{ }\text{
  }\mathrm{\ensuremath{\mu}}\mathrm{eV}$ mass at selected resolving powers},}\
  }\href {\doibase 10.1103/rlbt-65rc} {\bibfield  {journal} {\bibinfo
  {journal} {Phys. Rev. D}\ }\textbf {\bibinfo {volume} {112}},\ \bibinfo
  {pages} {L101101} (\bibinfo {year} {2025})}\BibitemShut {NoStop}%
\bibitem [{\citenamefont {Zhong}\ \emph {et~al.}(2018)\citenamefont {Zhong}
  \emph {et~al.}}]{HAYSTAC:2018rwy}%
  \BibitemOpen
  \bibfield  {author} {\bibinfo {author} {\bibfnamefont {L.}~\bibnamefont
  {Zhong}} \emph {et~al.} (\bibinfo {collaboration} {HAYSTAC}),\ }\bibfield
  {title} {\enquote {\bibinfo {title} {{Results from phase 1 of the HAYSTAC
  microwave cavity axion experiment}},}\ }\href {\doibase
  10.1103/PhysRevD.97.092001} {\bibfield  {journal} {\bibinfo  {journal} {Phys.
  Rev. D}\ }\textbf {\bibinfo {volume} {97}},\ \bibinfo {pages} {092001}
  (\bibinfo {year} {2018})},\ \Eprint {http://arxiv.org/abs/1803.03690}
  {arXiv:1803.03690 [hep-ex]} \BibitemShut {NoStop}%
\bibitem [{\citenamefont {Jewell}\ \emph {et~al.}(2023)\citenamefont {Jewell}
  \emph {et~al.}}]{HAYSTAC:2023cam}%
  \BibitemOpen
  \bibfield  {author} {\bibinfo {author} {\bibfnamefont {M.~J.}\ \bibnamefont
  {Jewell}} \emph {et~al.} (\bibinfo {collaboration} {HAYSTAC Collaboration}),\
  }\bibfield  {title} {\enquote {\bibinfo {title} {New results from haystac's
  phase ii operation with a squeezed state receiver},}\ }\href {\doibase
  10.1103/PhysRevD.107.072007} {\bibfield  {journal} {\bibinfo  {journal}
  {Phys. Rev. D}\ }\textbf {\bibinfo {volume} {107}},\ \bibinfo {pages}
  {072007} (\bibinfo {year} {2023})}\BibitemShut {NoStop}%
\bibitem [{\citenamefont {Bai}\ \emph {et~al.}(2025)\citenamefont {Bai} \emph
  {et~al.}}]{HAYSTAC:2024jch}%
  \BibitemOpen
  \bibfield  {author} {\bibinfo {author} {\bibfnamefont {Xiran}\ \bibnamefont
  {Bai}} \emph {et~al.} (\bibinfo {collaboration} {HAYSTAC}),\ }\bibfield
  {title} {\enquote {\bibinfo {title} {{Dark Matter Axion Search with HAYSTAC
  Phase II}},}\ }\href {\doibase 10.1103/PhysRevLett.134.151006} {\bibfield
  {journal} {\bibinfo  {journal} {Phys. Rev. Lett.}\ }\textbf {\bibinfo
  {volume} {134}},\ \bibinfo {pages} {151006} (\bibinfo {year} {2025})},\
  \Eprint {http://arxiv.org/abs/2409.08998} {arXiv:2409.08998 [hep-ex]}
  \BibitemShut {NoStop}%
\bibitem [{\citenamefont {Kim}\ \emph {et~al.}(2023)\citenamefont {Kim} \emph
  {et~al.}}]{Kim:2022hmg}%
  \BibitemOpen
  \bibfield  {author} {\bibinfo {author} {\bibfnamefont {Jinsu}\ \bibnamefont
  {Kim}} \emph {et~al.},\ }\bibfield  {title} {\enquote {\bibinfo {title}
  {{Near-Quantum-Noise Axion Dark Matter Search at CAPP around
  9.5\,\,\ensuremath{\mu}eV}},}\ }\href {\doibase
  10.1103/PhysRevLett.130.091602} {\bibfield  {journal} {\bibinfo  {journal}
  {Phys. Rev. Lett.}\ }\textbf {\bibinfo {volume} {130}},\ \bibinfo {pages}
  {091602} (\bibinfo {year} {2023})},\ \Eprint
  {http://arxiv.org/abs/2207.13597} {arXiv:2207.13597 [hep-ex]} \BibitemShut
  {NoStop}%
\bibitem [{\citenamefont {McAllister}\ \emph {et~al.}(2017)\citenamefont
  {McAllister}, \citenamefont {Flower}, \citenamefont {Ivanov}, \citenamefont
  {Goryachev}, \citenamefont {Bourhill},\ and\ \citenamefont
  {Tobar}}]{McAllister:2017lkb}%
  \BibitemOpen
  \bibfield  {author} {\bibinfo {author} {\bibfnamefont {Ben~T.}\ \bibnamefont
  {McAllister}}, \bibinfo {author} {\bibfnamefont {Graeme}\ \bibnamefont
  {Flower}}, \bibinfo {author} {\bibfnamefont {Eugene~N.}\ \bibnamefont
  {Ivanov}}, \bibinfo {author} {\bibfnamefont {Maxim}\ \bibnamefont
  {Goryachev}}, \bibinfo {author} {\bibfnamefont {Jeremy}\ \bibnamefont
  {Bourhill}}, \ and\ \bibinfo {author} {\bibfnamefont {Michael~E.}\
  \bibnamefont {Tobar}},\ }\bibfield  {title} {\enquote {\bibinfo {title} {{The
  ORGAN Experiment: An axion haloscope above 15 GHz}},}\ }\href {\doibase
  10.1016/j.dark.2017.09.010} {\bibfield  {journal} {\bibinfo  {journal} {Phys.
  Dark Univ.}\ }\textbf {\bibinfo {volume} {18}},\ \bibinfo {pages} {67--72}
  (\bibinfo {year} {2017})},\ \Eprint {http://arxiv.org/abs/1706.00209}
  {arXiv:1706.00209 [physics.ins-det]} \BibitemShut {NoStop}%
\bibitem [{\citenamefont {Quiskamp}\ \emph {et~al.}(2022)\citenamefont
  {Quiskamp}, \citenamefont {McAllister}, \citenamefont {Altin}, \citenamefont
  {Ivanov}, \citenamefont {Goryachev},\ and\ \citenamefont
  {Tobar}}]{Quiskamp:2022pks}%
  \BibitemOpen
  \bibfield  {author} {\bibinfo {author} {\bibfnamefont {Aaron~P.}\
  \bibnamefont {Quiskamp}}, \bibinfo {author} {\bibfnamefont {Ben~T.}\
  \bibnamefont {McAllister}}, \bibinfo {author} {\bibfnamefont {Paul}\
  \bibnamefont {Altin}}, \bibinfo {author} {\bibfnamefont {Eugene~N.}\
  \bibnamefont {Ivanov}}, \bibinfo {author} {\bibfnamefont {Maxim}\
  \bibnamefont {Goryachev}}, \ and\ \bibinfo {author} {\bibfnamefont
  {Michael~E.}\ \bibnamefont {Tobar}},\ }\bibfield  {title} {\enquote {\bibinfo
  {title} {{Direct search for dark matter axions excluding ALP cogenesis in the
  63- to 67-\ensuremath{\mu}eV range with the ORGAN experiment}},}\ }\href
  {\doibase 10.1126/sciadv.abq3765} {\bibfield  {journal} {\bibinfo  {journal}
  {Sci. Adv.}\ }\textbf {\bibinfo {volume} {8}},\ \bibinfo {pages} {abq3765}
  (\bibinfo {year} {2022})},\ \Eprint {http://arxiv.org/abs/2203.12152}
  {arXiv:2203.12152 [hep-ex]} \BibitemShut {NoStop}%
\bibitem [{\citenamefont {Quiskamp}\ \emph {et~al.}(2023)\citenamefont
  {Quiskamp}, \citenamefont {McAllister}, \citenamefont {Altin}, \citenamefont
  {Ivanov}, \citenamefont {Goryachev},\ and\ \citenamefont
  {Tobar}}]{Quiskamp:2023ehr}%
  \BibitemOpen
  \bibfield  {author} {\bibinfo {author} {\bibfnamefont {Aaron}\ \bibnamefont
  {Quiskamp}}, \bibinfo {author} {\bibfnamefont {Ben~T.}\ \bibnamefont
  {McAllister}}, \bibinfo {author} {\bibfnamefont {Paul}\ \bibnamefont
  {Altin}}, \bibinfo {author} {\bibfnamefont {Eugene~N.}\ \bibnamefont
  {Ivanov}}, \bibinfo {author} {\bibfnamefont {Maxim}\ \bibnamefont
  {Goryachev}}, \ and\ \bibinfo {author} {\bibfnamefont {Michael~E.}\
  \bibnamefont {Tobar}},\ }\bibfield  {title} {\enquote {\bibinfo {title}
  {{Exclusion of ALP Cogenesis Dark Matter in a Mass Window Above 100
  $\mu$eV}},}\ }\href@noop {} {\  (\bibinfo {year} {2023})},\ \Eprint
  {http://arxiv.org/abs/2310.00904} {arXiv:2310.00904 [hep-ex]} \BibitemShut
  {NoStop}%
\bibitem [{\citenamefont {Quiskamp}\ \emph {et~al.}(2024)\citenamefont
  {Quiskamp}, \citenamefont {Flower}, \citenamefont {Samuels}, \citenamefont
  {McAllister}, \citenamefont {Altin}, \citenamefont {Ivanov}, \citenamefont
  {Goryachev},\ and\ \citenamefont {Tobar}}]{Quiskamp:2024oet}%
  \BibitemOpen
  \bibfield  {author} {\bibinfo {author} {\bibfnamefont {Aaron~P.}\
  \bibnamefont {Quiskamp}}, \bibinfo {author} {\bibfnamefont {Graeme}\
  \bibnamefont {Flower}}, \bibinfo {author} {\bibfnamefont {Steven}\
  \bibnamefont {Samuels}}, \bibinfo {author} {\bibfnamefont {Ben~T.}\
  \bibnamefont {McAllister}}, \bibinfo {author} {\bibfnamefont {Paul}\
  \bibnamefont {Altin}}, \bibinfo {author} {\bibfnamefont {Eugene~N.}\
  \bibnamefont {Ivanov}}, \bibinfo {author} {\bibfnamefont {Maxim}\
  \bibnamefont {Goryachev}}, \ and\ \bibinfo {author} {\bibfnamefont
  {Michael~E.}\ \bibnamefont {Tobar}},\ }\bibfield  {title} {\enquote {\bibinfo
  {title} {{Near-quantum limited axion dark matter search with the ORGAN
  experiment around 26 $\mu$eV}},}\ }\href@noop {} {\  (\bibinfo {year}
  {2024})},\ \Eprint {http://arxiv.org/abs/2407.18586} {arXiv:2407.18586
  [hep-ex]} \BibitemShut {NoStop}%
\bibitem [{\citenamefont {Grenet}\ \emph {et~al.}(2021)\citenamefont {Grenet},
  \citenamefont {Ballou}, \citenamefont {Basto}, \citenamefont {Martineau},
  \citenamefont {Perrier}, \citenamefont {Pugnat}, \citenamefont {Quevillon},
  \citenamefont {Roch},\ and\ \citenamefont {Smith}}]{Grenet:2021vbb}%
  \BibitemOpen
  \bibfield  {author} {\bibinfo {author} {\bibfnamefont {Thierry}\ \bibnamefont
  {Grenet}}, \bibinfo {author} {\bibfnamefont {Rafik}\ \bibnamefont {Ballou}},
  \bibinfo {author} {\bibfnamefont {Quentin}\ \bibnamefont {Basto}}, \bibinfo
  {author} {\bibfnamefont {Killian}\ \bibnamefont {Martineau}}, \bibinfo
  {author} {\bibfnamefont {Pierre}\ \bibnamefont {Perrier}}, \bibinfo {author}
  {\bibfnamefont {Pierre}\ \bibnamefont {Pugnat}}, \bibinfo {author}
  {\bibfnamefont {J\'er\'emie}\ \bibnamefont {Quevillon}}, \bibinfo {author}
  {\bibfnamefont {Nicolas}\ \bibnamefont {Roch}}, \ and\ \bibinfo {author}
  {\bibfnamefont {Christopher}\ \bibnamefont {Smith}},\ }\bibfield  {title}
  {\enquote {\bibinfo {title} {{The Grenoble Axion Haloscope platform (GrAHal):
  development plan and first results}},}\ }\href@noop {} {\  (\bibinfo {year}
  {2021})},\ \Eprint {http://arxiv.org/abs/2110.14406} {arXiv:2110.14406
  [hep-ex]} \BibitemShut {NoStop}%
\bibitem [{\citenamefont {Adair}\ \emph {et~al.}(2022)\citenamefont {Adair}
  \emph {et~al.}}]{Adair:2022rtw}%
  \BibitemOpen
  \bibfield  {author} {\bibinfo {author} {\bibfnamefont {C.~M.}\ \bibnamefont
  {Adair}} \emph {et~al.},\ }\bibfield  {title} {\enquote {\bibinfo {title}
  {{Search for Dark Matter Axions with CAST-CAPP}},}\ }\href {\doibase
  10.1038/s41467-022-33913-6} {\bibfield  {journal} {\bibinfo  {journal}
  {Nature Commun.}\ }\textbf {\bibinfo {volume} {13}},\ \bibinfo {pages} {6180}
  (\bibinfo {year} {2022})},\ \Eprint {http://arxiv.org/abs/2211.02902}
  {arXiv:2211.02902 [hep-ex]} \BibitemShut {NoStop}%
\bibitem [{\citenamefont {Boutan}\ \emph {et~al.}(2018)\citenamefont {Boutan}
  \emph {et~al.}}]{ADMX:2018ogs}%
  \BibitemOpen
  \bibfield  {author} {\bibinfo {author} {\bibfnamefont {C.}~\bibnamefont
  {Boutan}} \emph {et~al.} (\bibinfo {collaboration} {ADMX}),\ }\bibfield
  {title} {\enquote {\bibinfo {title} {{Piezoelectrically Tuned Multimode
  Cavity Search for Axion Dark Matter}},}\ }\href {\doibase
  10.1103/PhysRevLett.121.261302} {\bibfield  {journal} {\bibinfo  {journal}
  {Phys. Rev. Lett.}\ }\textbf {\bibinfo {volume} {121}},\ \bibinfo {pages}
  {261302} (\bibinfo {year} {2018})},\ \Eprint
  {http://arxiv.org/abs/1901.00920} {arXiv:1901.00920 [hep-ex]} \BibitemShut
  {NoStop}%
\bibitem [{\citenamefont {Bartram}\ \emph {et~al.}(2021)\citenamefont {Bartram}
  \emph {et~al.}}]{Bartram:2021ysp}%
  \BibitemOpen
  \bibfield  {author} {\bibinfo {author} {\bibfnamefont {C.}~\bibnamefont
  {Bartram}} \emph {et~al.},\ }\bibfield  {title} {\enquote {\bibinfo {title}
  {{Dark Matter Axion Search Using a Josephson Traveling Wave Parametric
  Amplifier}},}\ }\href@noop {} {\  (\bibinfo {year} {2021})},\ \Eprint
  {http://arxiv.org/abs/2110.10262} {arXiv:2110.10262 [hep-ex]} \BibitemShut
  {NoStop}%
\bibitem [{\citenamefont {Hoshino}\ \emph {et~al.}(2025)\citenamefont {Hoshino}
  \emph {et~al.}}]{Hoshino:2025fiz}%
  \BibitemOpen
  \bibfield  {author} {\bibinfo {author} {\bibfnamefont {Gabe}\ \bibnamefont
  {Hoshino}} \emph {et~al.},\ }\bibfield  {title} {\enquote {\bibinfo {title}
  {{First Axion-Like Particle Results from a Broadband Search for Wave-Like
  Dark Matter in the 44 to 52 $\mu$eV Range with a Coaxial Dish Antenna}},}\
  }\href@noop {} {\  (\bibinfo {year} {2025})},\ \Eprint
  {http://arxiv.org/abs/2501.17119} {arXiv:2501.17119 [hep-ex]} \BibitemShut
  {NoStop}%
\bibitem [{\citenamefont {Alesini}\ \emph {et~al.}(2019)\citenamefont {Alesini}
  \emph {et~al.}}]{Alesini:2019ajt}%
  \BibitemOpen
  \bibfield  {author} {\bibinfo {author} {\bibfnamefont {D.}~\bibnamefont
  {Alesini}} \emph {et~al.},\ }\bibfield  {title} {\enquote {\bibinfo {title}
  {{Galactic axions search with a superconducting resonant cavity}},}\ }\href
  {\doibase 10.1103/PhysRevD.99.101101} {\bibfield  {journal} {\bibinfo
  {journal} {Phys. Rev. D}\ }\textbf {\bibinfo {volume} {99}},\ \bibinfo
  {pages} {101101} (\bibinfo {year} {2019})},\ \Eprint
  {http://arxiv.org/abs/1903.06547} {arXiv:1903.06547 [physics.ins-det]}
  \BibitemShut {NoStop}%
\bibitem [{\citenamefont {Alesini}\ \emph {et~al.}(2021)\citenamefont {Alesini}
  \emph {et~al.}}]{Alesini:2020vny}%
  \BibitemOpen
  \bibfield  {author} {\bibinfo {author} {\bibfnamefont {D.}~\bibnamefont
  {Alesini}} \emph {et~al.},\ }\bibfield  {title} {\enquote {\bibinfo {title}
  {{Search for invisible axion dark matter of mass m$_a=43~\mu$eV with the
  QUAX--$a\gamma$ experiment}},}\ }\href {\doibase 10.1103/PhysRevD.103.102004}
  {\bibfield  {journal} {\bibinfo  {journal} {Phys. Rev. D}\ }\textbf {\bibinfo
  {volume} {103}},\ \bibinfo {pages} {102004} (\bibinfo {year} {2021})},\
  \Eprint {http://arxiv.org/abs/2012.09498} {arXiv:2012.09498 [hep-ex]}
  \BibitemShut {NoStop}%
\bibitem [{\citenamefont {Alesini}\ \emph {et~al.}(2022)\citenamefont {Alesini}
  \emph {et~al.}}]{Alesini:2022lnp}%
  \BibitemOpen
  \bibfield  {author} {\bibinfo {author} {\bibfnamefont {D.}~\bibnamefont
  {Alesini}} \emph {et~al.},\ }\bibfield  {title} {\enquote {\bibinfo {title}
  {{Search for Galactic axions with a high-Q dielectric cavity}},}\ }\href
  {\doibase 10.1103/PhysRevD.106.052007} {\bibfield  {journal} {\bibinfo
  {journal} {Phys. Rev. D}\ }\textbf {\bibinfo {volume} {106}},\ \bibinfo
  {pages} {052007} (\bibinfo {year} {2022})},\ \Eprint
  {http://arxiv.org/abs/2208.12670} {arXiv:2208.12670 [hep-ex]} \BibitemShut
  {NoStop}%
\bibitem [{\citenamefont {Di~Vora}\ \emph {et~al.}(2023)\citenamefont {Di~Vora}
  \emph {et~al.}}]{QUAX:2023gop}%
  \BibitemOpen
  \bibfield  {author} {\bibinfo {author} {\bibfnamefont {R.}~\bibnamefont
  {Di~Vora}} \emph {et~al.} (\bibinfo {collaboration} {QUAX}),\ }\bibfield
  {title} {\enquote {\bibinfo {title} {{Search for galactic axions with a
  traveling wave parametric amplifier}},}\ }\href {\doibase
  10.1103/PhysRevD.108.062005} {\bibfield  {journal} {\bibinfo  {journal}
  {Phys. Rev. D}\ }\textbf {\bibinfo {volume} {108}},\ \bibinfo {pages}
  {062005} (\bibinfo {year} {2023})},\ \Eprint
  {http://arxiv.org/abs/2304.07505} {arXiv:2304.07505 [hep-ex]} \BibitemShut
  {NoStop}%
\bibitem [{\citenamefont {Rettaroli}\ \emph {et~al.}(2024)\citenamefont
  {Rettaroli} \emph {et~al.}}]{QUAX:2024fut}%
  \BibitemOpen
  \bibfield  {author} {\bibinfo {author} {\bibfnamefont {A.}~\bibnamefont
  {Rettaroli}} \emph {et~al.} (\bibinfo {collaboration} {QUAX}),\ }\bibfield
  {title} {\enquote {\bibinfo {title} {{Search for axion dark matter with the
  QUAX\textendash{}LNF tunable haloscope}},}\ }\href {\doibase
  10.1103/PhysRevD.110.022008} {\bibfield  {journal} {\bibinfo  {journal}
  {Phys. Rev. D}\ }\textbf {\bibinfo {volume} {110}},\ \bibinfo {pages}
  {022008} (\bibinfo {year} {2024})},\ \Eprint
  {http://arxiv.org/abs/2402.19063} {arXiv:2402.19063 [physics.ins-det]}
  \BibitemShut {NoStop}%
\bibitem [{\citenamefont {Melc\'on}\ \emph {et~al.}(2020)\citenamefont
  {Melc\'on} \emph {et~al.}}]{CAST:2020rlf}%
  \BibitemOpen
  \bibfield  {author} {\bibinfo {author} {\bibfnamefont {A.~\'Alvarez}\
  \bibnamefont {Melc\'on}} \emph {et~al.} (\bibinfo {collaboration} {CAST}),\
  }\bibfield  {title} {\enquote {\bibinfo {title} {{First results of the
  CAST-RADES haloscope search for axions at 34.67 $\mu$eV}},}\ }\href {\doibase
  10.1007/JHEP10(2021)075} {\bibfield  {journal} {\bibinfo  {journal} {JHEP}\
  }\textbf {\bibinfo {volume} {21}},\ \bibinfo {pages} {075} (\bibinfo {year}
  {2020})},\ \Eprint {http://arxiv.org/abs/2104.13798} {arXiv:2104.13798
  [hep-ex]} \BibitemShut {NoStop}%
\bibitem [{\citenamefont {Ahyoune}\ \emph {et~al.}(2024)\citenamefont {Ahyoune}
  \emph {et~al.}}]{Ahyoune:2024klt}%
  \BibitemOpen
  \bibfield  {author} {\bibinfo {author} {\bibfnamefont {S.}~\bibnamefont
  {Ahyoune}} \emph {et~al.},\ }\bibfield  {title} {\enquote {\bibinfo {title}
  {{RADES axion search results with a High-Temperature Superconducting cavity
  in an 11.7 T magnet}},}\ }\href@noop {} {\  (\bibinfo {year} {2024})},\
  \Eprint {http://arxiv.org/abs/2403.07790} {arXiv:2403.07790 [hep-ex]}
  \BibitemShut {NoStop}%
\bibitem [{\citenamefont {Garcia}\ \emph {et~al.}(2024)\citenamefont {Garcia}
  \emph {et~al.}}]{Garcia:2024xzc}%
  \BibitemOpen
  \bibfield  {author} {\bibinfo {author} {\bibfnamefont {B.~Ary dos~Santos}\
  \bibnamefont {Garcia}} \emph {et~al.},\ }\bibfield  {title} {\enquote
  {\bibinfo {title} {{First search for axion dark matter with a Madmax
  prototype}},}\ }\href@noop {} {\  (\bibinfo {year} {2024})},\ \Eprint
  {http://arxiv.org/abs/2409.11777} {arXiv:2409.11777 [hep-ex]} \BibitemShut
  {NoStop}%
\bibitem [{\citenamefont {Padamsee}\ \emph {et~al.}(John Wiley \& Sons, New
  York, 1998)\citenamefont {Padamsee}, \citenamefont {Knobloch},\ and\
  \citenamefont {Hays}}]{Padamsee:1998vf}%
  \BibitemOpen
  \bibfield  {author} {\bibinfo {author} {\bibfnamefont {H.}~\bibnamefont
  {Padamsee}}, \bibinfo {author} {\bibfnamefont {J.}~\bibnamefont {Knobloch}},
  \ and\ \bibinfo {author} {\bibfnamefont {T.}~\bibnamefont {Hays}},\
  }\href@noop {} {\emph {\bibinfo {title} {{RF superconductivity for
  accelerators}}}}\ (\bibinfo {year} {John Wiley \& Sons, New York,
  1998})\BibitemShut {NoStop}%
\bibitem [{\citenamefont {Ouellet}\ \emph {et~al.}(2019)\citenamefont {Ouellet}
  \emph {et~al.}}]{Ouellet:2018beu}%
  \BibitemOpen
  \bibfield  {author} {\bibinfo {author} {\bibfnamefont {Jonathan~L.}\
  \bibnamefont {Ouellet}} \emph {et~al.},\ }\bibfield  {title} {\enquote
  {\bibinfo {title} {{First Results from ABRACADABRA-10 cm: A Search for
  Sub-$\mu$eV Axion Dark Matter}},}\ }\href {\doibase
  10.1103/PhysRevLett.122.121802} {\bibfield  {journal} {\bibinfo  {journal}
  {Phys. Rev. Lett.}\ }\textbf {\bibinfo {volume} {122}},\ \bibinfo {pages}
  {121802} (\bibinfo {year} {2019})},\ \Eprint
  {http://arxiv.org/abs/1810.12257} {arXiv:1810.12257 [hep-ex]} \BibitemShut
  {NoStop}%
\bibitem [{\citenamefont {Gramolin}\ \emph {et~al.}(2021)\citenamefont
  {Gramolin}, \citenamefont {Aybas}, \citenamefont {Johnson}, \citenamefont
  {Adam},\ and\ \citenamefont {Sushkov}}]{Gramolin:2020ict}%
  \BibitemOpen
  \bibfield  {author} {\bibinfo {author} {\bibfnamefont {Alexander~V.}\
  \bibnamefont {Gramolin}}, \bibinfo {author} {\bibfnamefont {Deniz}\
  \bibnamefont {Aybas}}, \bibinfo {author} {\bibfnamefont {Dorian}\
  \bibnamefont {Johnson}}, \bibinfo {author} {\bibfnamefont {Janos}\
  \bibnamefont {Adam}}, \ and\ \bibinfo {author} {\bibfnamefont {Alexander~O.}\
  \bibnamefont {Sushkov}},\ }\bibfield  {title} {\enquote {\bibinfo {title}
  {{Search for axion-like dark matter with ferromagnets}},}\ }\href {\doibase
  10.1038/s41567-020-1006-6} {\bibfield  {journal} {\bibinfo  {journal} {Nature
  Phys.}\ }\textbf {\bibinfo {volume} {17}},\ \bibinfo {pages} {79--84}
  (\bibinfo {year} {2021})},\ \Eprint {http://arxiv.org/abs/2003.03348}
  {arXiv:2003.03348 [hep-ex]} \BibitemShut {NoStop}%
\bibitem [{\citenamefont {Salemi}\ \emph {et~al.}(2021)\citenamefont {Salemi}
  \emph {et~al.}}]{Salemi:2021gck}%
  \BibitemOpen
  \bibfield  {author} {\bibinfo {author} {\bibfnamefont {Chiara~P.}\
  \bibnamefont {Salemi}} \emph {et~al.},\ }\bibfield  {title} {\enquote
  {\bibinfo {title} {{Search for Low-Mass Axion Dark Matter with
  ABRACADABRA-10~cm}},}\ }\href {\doibase 10.1103/PhysRevLett.127.081801}
  {\bibfield  {journal} {\bibinfo  {journal} {Phys. Rev. Lett.}\ }\textbf
  {\bibinfo {volume} {127}},\ \bibinfo {pages} {081801} (\bibinfo {year}
  {2021})},\ \Eprint {http://arxiv.org/abs/2102.06722} {arXiv:2102.06722
  [hep-ex]} \BibitemShut {NoStop}%
\bibitem [{\citenamefont {Heinze}\ \emph {et~al.}(2023)\citenamefont {Heinze},
  \citenamefont {Gill}, \citenamefont {Dmitriev}, \citenamefont {Smetana},
  \citenamefont {Yan}, \citenamefont {Boyer}, \citenamefont {Martynov},\ and\
  \citenamefont {Evans}}]{Heinze:2023nfb}%
  \BibitemOpen
  \bibfield  {author} {\bibinfo {author} {\bibfnamefont {Joscha}\ \bibnamefont
  {Heinze}}, \bibinfo {author} {\bibfnamefont {Alex}\ \bibnamefont {Gill}},
  \bibinfo {author} {\bibfnamefont {Artemiy}\ \bibnamefont {Dmitriev}},
  \bibinfo {author} {\bibfnamefont {Jiri}\ \bibnamefont {Smetana}}, \bibinfo
  {author} {\bibfnamefont {Tiangliang}\ \bibnamefont {Yan}}, \bibinfo {author}
  {\bibfnamefont {Vincent}\ \bibnamefont {Boyer}}, \bibinfo {author}
  {\bibfnamefont {Denis}\ \bibnamefont {Martynov}}, \ and\ \bibinfo {author}
  {\bibfnamefont {Matthew}\ \bibnamefont {Evans}},\ }\bibfield  {title}
  {\enquote {\bibinfo {title} {{First results of the Laser-Interferometric
  Detector for Axions (LIDA)}},}\ }\href@noop {} {\  (\bibinfo {year}
  {2023})},\ \Eprint {http://arxiv.org/abs/2307.01365} {arXiv:2307.01365
  [astro-ph.CO]} \BibitemShut {NoStop}%
\bibitem [{\citenamefont {Brubaker}\ \emph
  {et~al.}(2017{\natexlab{b}})\citenamefont {Brubaker}, \citenamefont {Zhong},
  \citenamefont {Lamoreaux}, \citenamefont {Lehnert},\ and\ \citenamefont {van
  Bibber}}]{Brubaker:2017rna}%
  \BibitemOpen
  \bibfield  {author} {\bibinfo {author} {\bibfnamefont {B.~M.}\ \bibnamefont
  {Brubaker}}, \bibinfo {author} {\bibfnamefont {L.}~\bibnamefont {Zhong}},
  \bibinfo {author} {\bibfnamefont {S.~K.}\ \bibnamefont {Lamoreaux}}, \bibinfo
  {author} {\bibfnamefont {K.~W.}\ \bibnamefont {Lehnert}}, \ and\ \bibinfo
  {author} {\bibfnamefont {K.~A.}\ \bibnamefont {van Bibber}},\ }\bibfield
  {title} {\enquote {\bibinfo {title} {{HAYSTAC axion search analysis
  procedure}},}\ }\href {\doibase 10.1103/PhysRevD.96.123008} {\bibfield
  {journal} {\bibinfo  {journal} {Phys. Rev. D}\ }\textbf {\bibinfo {volume}
  {96}},\ \bibinfo {pages} {123008} (\bibinfo {year} {2017}{\natexlab{b}})},\
  \Eprint {http://arxiv.org/abs/1706.08388} {arXiv:1706.08388 [astro-ph.IM]}
  \BibitemShut {NoStop}%
\end{thebibliography}%
